\documentclass[
  aps,pra,reprint,superscriptaddress,
  amsmath,amssymb,longbibliography
]{revtex4-2}

\usepackage{graphicx}
\usepackage{bm}
\usepackage{booktabs}
\usepackage[hidelinks]{hyperref}

\graphicspath{{figures/}}

\DeclareMathOperator{\Tr}{Tr}
\DeclareMathOperator{\rank}{rank}
\DeclareMathOperator{\Sym}{Sym}
\DeclareMathOperator{\spanop}{span}
\newcommand{\ket}[1]{\lvert #1\rangle}
\newcommand{\bra}[1]{\langle #1\rvert}
\newcommand{\braket}[2]{\langle #1\vert #2\rangle}
\newcommand{\Hilbert}{\mathcal H}

\newcommand{\pcert}{p_{\mathrm{cert}}}
\newcommand{\pbin}{p_{\mathrm{bin}}}
\newtheorem{theorem}{Theorem}

\newtheorem{corollary}[theorem]{Corollary}

\newif\ifwithbibliography
\withbibliographytrue

\begin{document}

\title{Optimal finite-copy readout of three-qubit entanglement with state retention}

\author{Yu-Xin Wu}
\affiliation{School of Science, Zhejiang University of Science and Technology, Hangzhou 310023, China}
\author{Yuan-Hong Tao}
\thanks{Corresponding author: Yuanhong Tao, taoyuanhong12@126.com}
\affiliation{School of Science, Zhejiang University of Science and Technology, Hangzhou 310023, China}
\author{Ji-Hui Mo}
\affiliation{School of Science, Zhejiang University of Science and Technology, Hangzhou 310023, China}
\author{Peng-Fei Ma}
\affiliation{School of Science, Zhejiang University of Science and Technology, Hangzhou 310023, China}
\author{Shu-Hui Wu}
\affiliation{School of Science, Zhejiang University of Science and Technology, Hangzhou 310023, China}
\affiliation{ZUST-CYU City of the Future Engineering Institute, Zhejiang University of Science and Technology, Hangzhou 310023, China}
\author{Shao-Ming Fei}
\affiliation{School of Mathematical Sciences, Capital Normal University, Beijing 100048, China}

\begin{abstract}
For unknown three-qubit pure states, we characterize the optimal trade-off among GHZ-class certification, binary probability readout of the three-tangle, and conditional state retention. The vanishing ideal of the null class specifies the optimal certification subspace, while a polynomial multiplication operator gives the spectral limit for binary readout. Four copies are necessary for a nonzero
certificate. A fifth copy raises the optimal success probability from
$\tau^2/48$ to $5\tau^2/72$ and permits the original unknown state to be
retained, where $\tau$ is the three-tangle. We obtain this gain as an instance
of a general symmetry-based additional-copy theorem. At six copies the optimal
certification probability is $(39+2S)\tau^2/296$, with
$S=\sum_X\operatorname{Tr}\rho_X^2$, whereas the largest exact binary
probability is $\tau^2/8$. Local Schur measurements and conditional local
operations with classical communication attain these bounds. Retaining the
full six-copy branch record yields a weighted estimator with lower variance
than binary thinning of the same data. Explicit circuits and fixed-budget
comparisons connect the success probabilities to input-copy consumption,
retained quantum output, and coherent gate cost.
\end{abstract}

\maketitle

\section{Introduction}
\label{sec:introduction}

Measuring an entanglement invariant, certifying an entanglement class, and
leaving an entangled state available for subsequent use are different physical
tasks. A collective measurement on several identical preparations can accomplish
more than a measurement on one preparation, but it also consumes a larger
quantum input. The relevant resource question is therefore not only whether an
invariant can be measured. It is how efficiently a fixed measurement converts
copies of an unknown state into a conclusive classical outcome, and what quantum
state remains after that outcome.

Three-qubit pure states provide a setting in which these questions can be
answered explicitly. The three-tangle distinguishes the GHZ class from its
algebraic boundary, which includes the W class as well as biseparable and product
states~\cite{coffman2000,dur2000}. Polynomial invariants can be evaluated through
collective observables on repeated preparations~\cite{leifer2004,brun2004}.
Local projections onto symmetric and antisymmetric sectors also underlie
concurrence measurements~\cite{mintert2005,walborn2006}. For the three-tangle,
Yu and Song constructed a four-copy product of local singlet
projections~\cite{yu2007}, and Sugita developed a systematic construction using
local irreducible representations~\cite{sugita2008}. More recent proposals give
explicit direct-readout circuits~\cite{yahyavi2022,liu2026} or learn a local
canonical form before estimating the tangle~\cite{perezsalinas2020}.
A two-copy construction instead adjusts a measurement basis and reconstructs the
invariant from several probabilities~\cite{yu2016}. These approaches provide
different combinations of coherent block size, classical processing, and
measurement output.

Local Schur decompositions provide a second relevant connection. Universal
entanglement concentration uses them to separate representation degrees of
freedom from entangled multiplicity registers~\cite{matsumoto2007}.
Finite-copy limits have also been obtained for universal conclusive
Bell-pair concentration using concatenated two-qubit
operations~\cite{orthey2026}.
Multipartite versions reveal fixed Kronecker states and their relation to
covariants~\cite{botero2018,gonzalez2025}. In particular, the relevant five-copy
multiplicity space is one-dimensional, whereas a six-copy sector has two
independent directions~\cite{gonzalez2025}. These structures suggest collective
local measurements, but their operational value depends on which success
effect is required and which degrees of freedom must survive.

Here we characterize the optimal success effect for error-free certification
against a homogeneous algebraic null set. For a principal vanishing ideal,
polynomial multiplication gives both the certification projector and a spectral
bound on exact binary probability readout. A symmetry argument then identifies
a universal gain from one additional copy and a success map that returns the
unknown input state. Applied to three qubits, this construction gives exact
four-, five-, and six-copy limits and local protocols attaining them.
Five copies are necessary and sufficient for conclusive GHZ-class certification
that retains one unknown input state. Six copies exhibit a different effect:
optimal certification depends on both the three-tangle and local purities,
whereas optimal binary readout has a state-independent proportionality constant.
The distinction can be implemented by classical processing of the same local
measurement record.

We connect these results to fixed-input-budget statistics and explicit local
circuits. The five-copy protocol is compared with measuring four copies while
reserving a fifth, and the six-copy record is used to construct a weighted
estimator. Circuit resources expose the additional coherent processing needed
to obtain the copy advantage. Together, these comparisons identify the
information and retained quantum output supplied by each extra copy.

\section{Tasks and main results}
\label{sec:tasks}

Let $\Hilbert=(\mathbb C^2)^{\otimes3}$, and let each batch consist of
$n$ identical copies of an unknown normalized state $\ket\psi\in\Hilbert$.
Alice, Bob, and Charlie each hold their respective $n$ qubits. We first
optimize over all collective measurements, and then construct implementations
using operations within each party and classical communication (LOCC).
The measurement is fixed independently of $\psi$.

For Cayley's quartic hyperdeterminant $D$, define
\begin{equation}
 \tau=4|D|,\qquad t=\tau^2,\qquad
 S=\sum_{X=A,B,C}\Tr\rho_X^2.
 \label{eq:invariants}
\end{equation}
Here $\tau$ is the CKW three-tangle~\cite{coffman2000}; its values on normalized
pure states lie in $[0,1]$. The null hypothesis is the set $D(\psi)=0$.
For three-qubit pure states, its complement is precisely the GHZ SLOCC
class~\cite{dur2000}.

A success effect $0\le M\le I$ gives probability
$p_M(\psi)=\bra\psi^{\otimes n}M\ket\psi^{\otimes n}$.
We distinguish three requirements:
\begin{enumerate}
 \item \emph{Error-free certification:} $p_M(\phi)=0$ for every null state
 $D(\phi)=0$. A success certifies the GHZ class; the other outcome is
 inconclusive.
 \item \emph{Exact binary probability readout:} $p_M(\psi)=\kappa t(\psi)$
 for every normalized pure input, with a constant $\kappa$ to be maximized.
 Exactness refers to the probability identity.
 \item \emph{Conditional state retention:} the success operation returns an
 output register with normalized state $\ket\psi\bra\psi$.
\end{enumerate}
Retention is imposed in addition to certification. Failure branches can destroy
the input. The third requirement concerns the quantum instrument, whereas the
first two concern its success effect.

Only the compression of $M$ to
$\Hilbert_n=\Sym^n\Hilbert$ affects the allowed inputs. All effect inequalities
and uniqueness statements below refer to this compression. A physical local
instrument also specifies an action on the full tensor-product space; that
action determines its response to preparation errors.

\begin{figure*}[t]
 \includegraphics[width=\textwidth]{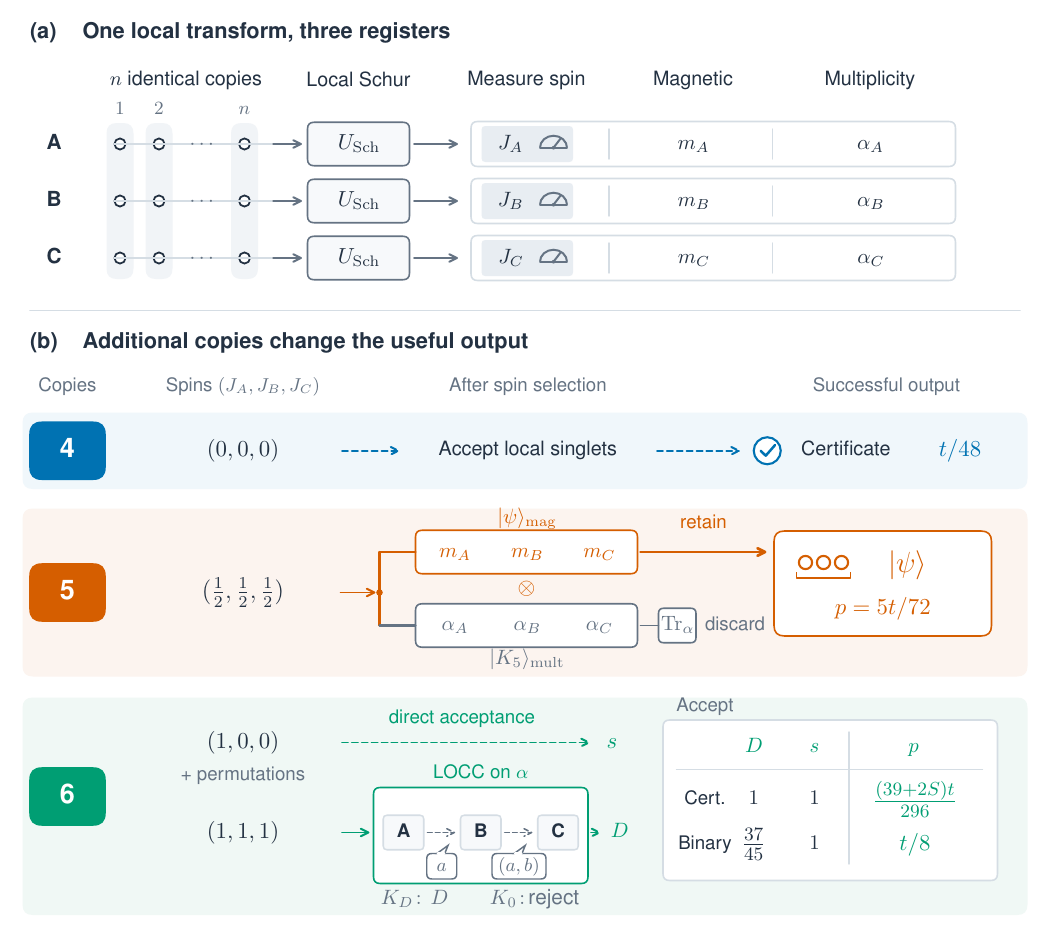}
 \caption{Local register structure and the output gained from additional
 copies. (a) Each column of the input array belongs to one copy of the
 unknown three-qubit state $\ket\psi$; each row belongs to one party.
 A local Schur transform separates total spin $J$, magnetic label $m$,
 and multiplicity label $\alpha$, which remain correlated in general.
 Only $J$, marked by a measurement dial, is measured at this stage.
 (b) The accepted spin sectors determine the subsequent operation.
 Four copies give a certificate. At five copies the three magnetic
 registers jointly carry $\ket\psi$ (the grouped qubit symbols), while
 the three multiplicity registers factor into $\ket{K_5}$ and are traced out.
 At six copies the directly
 accepted spin sectors form branch $s$. In the $(1,1,1)$ sector,
 Alice, Bob, and Charlie distinguish $K_D$ from $K_0$ by conditional
 measurements of $\alpha$, leaving $m$ as a spectator. Message bubbles
 label the communicated outcomes $a$ and $(a,b)$. The table gives
 the final acceptance weights for branches $D$ and $s$, followed by the
 resulting success probability: certification keeps both branches;
 binary readout thins $D$ by $37/45$. Solid paths carry quantum registers;
 dashed paths denote classical communication or acceptance. Other spin
 outcomes and $K_0$ are inconclusive. Here $t=\tau^2$ and
 $S=\sum_X\Tr\rho_X^2$.}
 \label{fig:protocol}
\end{figure*}

Table~\ref{tab:main} states the exact optima. The four-copy local singlet
construction follows Sugita~\cite{sugita2008}; the certification bound proves
its optimality within the present task. The fifth copy raises the success
coefficient by $10/3$ and supplies an intact three-qubit output. At six copies,
certification accepts more events than an optimal binary probability readout
for every $t>0$:
\begin{equation}
 \pcert^{(6)}-\pbin^{(6)}=\frac{(1+S)t}{148}>0.
 \label{eq:task-gap}
\end{equation}
The protocol structure is shown in Fig.~\ref{fig:protocol}.

\begin{table}[t]
 \caption{Optimal success probabilities on identical three-qubit pure-state
 inputs. The five-copy protocol also returns one copy of the unknown state.
 The final column lists the retained output of these constructions.}
 \label{tab:main}
 \begin{ruledtabular}
 \begin{tabular}{c c c l}
 $n$ & $\pcert^{(n)}$ & $\pbin^{(n)}$ & Retained output\\
 \hline
 $<4$ & $0$ & $0$ & ---\\
 $4$ & $t/48$ & $t/48$ & ---\\
 $5$ & $5t/72$ & $5t/72$ & $\ket\psi$\\
 $6$ & $(39+2S)t/296$ & $t/8$ & ---
 \end{tabular}
 \end{ruledtabular}
\end{table}

The copy threshold concerns a fixed, conclusive event. Estimation from several
measurement settings, including the two-copy reconstruction of
Ref.~\cite{yu2016}, uses a different output statistic. We compare such tasks
through their declared input budgets in Sec.~\ref{sec:statistics}.

\section{Optimal certification and probability readout}
\label{sec:optimality}

\subsection{Certification from the vanishing ideal}

The null states determine a subspace that every successful certification
event must annihilate. For a complex homogeneous algebraic cone
$Z\subseteq\Hilbert$, define
\begin{equation}
 \mathcal K_n=\spanop\{\ket{\phi}^{\otimes n}:\phi\in Z\}
 \subseteq\Hilbert_n,\qquad \Hilbert_n=\Sym^n\Hilbert.
 \label{alg:null-space}
\end{equation}
Let $\Pi_n$ project onto $\Hilbert_n$. Throughout this section, orthogonal
complements and identity operators are taken within $\Hilbert_n$.

\begin{theorem}[Optimal certification]
\label{alg:certification}
Every effect $0\le M\le I$ with zero acceptance probability on $Z$ obeys
\begin{equation}
 0\le\Pi_nM\Pi_n\le P_n,\qquad P_n=\Pi_{\mathcal K_n^\perp}.
 \label{alg:projector-bound}
\end{equation}
The projector $P_n$ simultaneously maximizes the acceptance probability
for every input. A nonzero certification event exists precisely when
$I(Z)_n\ne\{0\}$, where $I(Z)_n$ is the degree-$n$ part of the vanishing
ideal of $Z$.
\end{theorem}

To prove the operator bound, positivity gives
$\|M^{1/2}\ket{\phi}^{\otimes n}\|^2=0$ for every null input.
Consequently $M$ annihilates $\mathcal K_n$, and its compression is a
positive contraction supported on $\mathcal K_n^\perp$.
Conversely, the projector onto that complement is a valid success
effect. The correspondence
$\ket{v}\mapsto\bra{v}\ket{z}^{\otimes n}$ identifies
$\mathcal K_n^\perp$, antilinearly, with the homogeneous polynomials
vanishing on $Z$. Thus the least admissible copy number is
\begin{equation}
 n_{\min}=\min\{n:I(Z)_n\ne\{0\}\}.
 \label{alg:minimum-degree}
\end{equation}

For GHZ-class certification, $Z=\{D=0\}$. The hyperdeterminant $D$ is
irreducible over $\mathbb C$ (Appendix~\ref{app:algebra}), so its
vanishing ideal is the principal ideal $(D)$. Since $D$ has degree four,
\begin{equation}
 n_{\min}=4,\qquad
 \rank P_n=\binom{n+3}{7}\quad(n\ge4).
 \label{alg:ghz-rank}
\end{equation}
In particular, the optimal success spaces have dimensions $1$, $8$,
and $36$ at four, five, and six copies. These bounds allow arbitrary
global measurements; local protocols attaining them are constructed in
Sec.~\ref{sec:local}.

\subsection{The multiplication operator and its spectrum}

Polynomial functions of a state can be measured on multiple
copies~\cite{leifer2004,brun2004}. Here their multiplication operators
also determine the largest coefficient permitted for a binary event.
Let $f(z)=\sum_{|\alpha|=m}c_\alpha z^\alpha$ be a nonzero homogeneous
polynomial on $\mathbb C^d$. In the normalized occupation basis,
\begin{equation}
 \ket{z}^{\otimes n}
 =\sum_{|\gamma|=n}\sqrt{\binom n\gamma}\,z^\gamma\ket{\gamma},
 \qquad \binom n\gamma=\frac{n!}{\prod_j\gamma_j!}.
 \label{alg:occupation}
\end{equation}
For $n\ge m$, define $V_n:\Sym^{n-m}\Hilbert\to\Sym^n\Hilbert$ by
\begin{equation}
 (V_n)_{\alpha+\beta,\beta}
 =\overline{c_\alpha}
 \sqrt{\frac{\binom{n-m}{\beta}}{\binom n{\alpha+\beta}}},
 \qquad B_n=V_n^\dagger V_n.
 \label{alg:multiplication}
\end{equation}
Multiplication by $f$ is injective, hence $B_n>0$. Direct substitution
in Eq.~\eqref{alg:occupation} gives the identity
\begin{equation}
 V_n^\dagger\ket{z}^{\otimes n}
 =f(z)\ket{z}^{\otimes(n-m)}.
 \label{alg:contraction}
\end{equation}
If $I(Z)=(f)$, the optimal certification projector and its probability
are therefore
\begin{align}
 P_n&=V_nB_n^{-1}V_n^\dagger,\label{alg:gram-projector}\\
 \pcert(\psi)
 &=|f(\psi)|^2
 \bra{\psi}^{\otimes(n-m)}B_n^{-1}
 \ket{\psi}^{\otimes(n-m)}.
 \label{alg:cert-probability}
\end{align}

\begin{theorem}[Exact binary readout]
\label{alg:readout}
For any nonzero homogeneous $f$ and $n\ge m$, the largest coefficient
$c_n$ in a fixed success probability $p(\psi)=c_n|f(\psi)|^2$, valid
for every normalized pure input, is
\begin{equation}
 c_n^{\mathrm{opt}}=\frac{1}{\lambda_{\max}(B_n)}.
 \label{alg:general-binary-bound}
\end{equation}
The unique optimal compressed effect is
$V_nV_n^\dagger/\lambda_{\max}(B_n)$.
\end{theorem}

Indeed, Eq.~\eqref{alg:contraction} supplies one operator with the
required probabilities. Equality of probabilities on all pure tensor
powers fixes a Hermitian operator on $\Hilbert_n$: expanding its
expectation in $z$ and $\bar z$ fixes every matrix element
(Appendix~\ref{app:algebra}). The required compressed effect must thus
be $c_nV_nV_n^\dagger$. Its largest eigenvalue cannot exceed one,
and the stated coefficient saturates this condition.

For $f=D$ and $t=16|D|^2$, the optimal coefficient in $p=\kappa_n t$ is
\begin{equation}
 \kappa_n^{\mathrm{opt}}=
 \frac{1}{16\lambda_{\max}(B_n)}.
 \label{alg:tangle-binary-bound}
\end{equation}
Certification and binary readout use the same optimal effect exactly
when $B_n$ is a scalar operator. Otherwise the binary effect attenuates
some directions in the certification space. This distinction concerns
the binary probability constraint; a measurement with several recorded
outcomes can retain further statistical information.

\subsection{An additional copy and state retention}

An irreducible symmetry fixes the improvement obtained from one
additional copy. Define
\begin{equation}
 q_f=\sum_{|\alpha|=m}
       \frac{|c_\alpha|^2}{\binom m\alpha}.
 \label{alg:polynomial-norm}
\end{equation}

\begin{theorem}[Additional-copy gain]
\label{alg:extra-copy}
Suppose a compact group acts irreducibly and unitarily on
$\Hilbert\cong\mathbb C^d$, and
$f(Uz)=\chi(U)f(z)$ with $|\chi(U)|=1$.
Then
\begin{equation}
 B_m=q_f,\qquad B_{m+1}=bI,\qquad
 b=\frac{d+m}{d(m+1)}q_f.
 \label{alg:extra-gram}
\end{equation}
The optimal binary success probabilities consequently satisfy
\begin{equation}
 \frac{p_{m+1}^{\mathrm{opt}}}{p_m^{\mathrm{opt}}}
 =\frac{d(m+1)}{d+m}\qquad(f(\psi)\ne0).
 \label{alg:extra-gain}
\end{equation}
The $(m+1)$-copy optimum is attained by the success Kraus operator
\begin{equation}
 K=\frac{V_{m+1}^\dagger}{\sqrt b},\qquad
 K\ket{\psi}^{\otimes(m+1)}
 =\frac{f(\psi)}{\sqrt b}\ket{\psi}.
 \label{alg:retention-kraus}
\end{equation}
If $I(Z)=(f)$, both effects are also optimal for certification.
\end{theorem}

The covariance of $V_{m+1}$ implies that $B_{m+1}$ commutes with the
input representation. Schur's lemma makes it a scalar, whose trace is
$(d+m)q_f/(m+1)$ by the bosonic creation-operator identity derived in
Appendix~\ref{app:algebra}. Moreover,
$K^\dagger K=V_{m+1}V_{m+1}^\dagger/b$ is a projector.
Equation~\eqref{alg:retention-kraus} therefore describes a physical
success operation returning the unknown input state.

For three qubits, the product representation of $\mathrm{SU}(2)^3$
is irreducible, $D$ is invariant, and $d=8$, $m=4$, $q_D=3$.
Consequently $B_4=3$ and $B_5=(9/10)I$, giving probabilities
$t/48$ and $5t/72$. The fifth copy increases the success probability
by $10/3$ while enabling state retention. Sec.~\ref{sec:local}
implements this operation with local collective measurements.

\begin{corollary}[Minimum retention cost]
\label{alg:retention-minimum}
Five copies are necessary and sufficient for a nonzero unambiguous
GHZ-class certificate that returns the unknown input state upon success
for every GHZ-class pure input.
\end{corollary}

Fewer than four copies give no certificate. At four copies the
certification space has rank one, so every successful Kraus operator
maps that space to an input-independent output vector. The normalized
successful output is therefore fixed, including when several successful
outcomes and their classical records are retained. It cannot equal
every unknown GHZ-class input. Equation~\eqref{alg:retention-kraus}
at five copies attains the required bound.

\section{Local protocols and state retention}
\label{sec:local}

The four-, five-, and six-copy optima can all be attained by local
operations and classical communication. Each party first resolves the
total spin of its own qubits. The distinction between the protocols lies
in the information retained after this measurement: a success flag for
four copies, an undisturbed quantum state for five copies, and a
conditional multiplicity measurement for six copies. A schematic summary of these protocols and their outputs is presented in Figure~\ref{fig:protocol}.

For one party, the Schur decomposition is
\begin{equation}
 (\mathbb C^2)^{\otimes n}
 =\bigoplus_J\mathcal U_J\otimes\mathcal V_{(n/2+J,n/2-J)},
 \label{loc:schur}
\end{equation}
where $\mathcal U_J$ carries spin $J$ and $\mathcal V_\lambda$ is the
multiplicity space. We call their basis labels $m$ and $\alpha$,
respectively. Schur transforms can preserve both registers coherently
while measuring only $J$~\cite{bacon2006,kirby2018}. The simultaneous
permutation symmetry of identical copies constrains the three
multiplicity registers to the invariant subspace of
$\mathcal V_{\lambda_A}\otimes\mathcal V_{\lambda_B}\otimes
\mathcal V_{\lambda_C}$. Such invariant vectors, often called Kronecker
states, underlie local entanglement concentration
~\cite{botero2018,gonzalez2025}.

\subsection{Four-copy certification}

The four-copy protocol accepts precisely when all three parties obtain
$J=0$. This is the full local singlet projection of
Ref.~\cite{sugita2008}. The local projector, in terms of
$\mathbf J=\frac12\sum_{r=1}^n\boldsymbol\sigma_r$, is
\begin{equation}
 Q_0^{(4)}=
 \frac{(\mathbf J^2-2I)(\mathbf J^2-6I)}{12}.
 \label{loc:q4}
\end{equation}
The invariant multiplicity is
$g((2,2),(2,2),(2,2))=1$. Consequently, the range of
$(Q_0^{(4)})^{\otimes3}$ within $\Sym^4\Hilbert$ is the single
direction representing $D$. Its compressed effect is $P_4$, and
\begin{equation}
 p_4=\frac{|D|^2}{3}=\frac{t}{48}.
 \label{loc:p4}
\end{equation}
Thus this local measurement attains both the certification bound of
Theorem~\ref{alg:certification} and the binary-readout bound of
Theorem~\ref{alg:readout}. Appendix~\ref{app:schur} gives the
multiplicity calculation and its relation to the optimal subspace.

\subsection{Five copies with a retained state}

With five copies, each party accepts the spin-$1/2$ sector and retains its
magnetic register. The local success projector is
\begin{equation}
 Q_{1/2}^{(5)}=
 \frac{(\mathbf J^2-\tfrac{15}{4}I)
             (\mathbf J^2-\tfrac{35}{4}I)}{24}.
 \label{loc:q5}
\end{equation}
The triple multiplicity space has a unique invariant vector
$\ket{K_5}\in(\mathbb C^5)^{\otimes3}$, with
$g((3,2),(3,2),(3,2))=1$~\cite{gonzalez2025}. This sector therefore
contains one copy of the original eight-dimensional local-unitary
representation. It equals the range of $V_5$, so the compressed local
success effect is $P_5$.

The unnormalized successful Schur-decoded state is
\begin{equation}
 \ket{\Psi_{5,\mathrm{succ}}}
 =\sqrt{\frac{10}{9}}\,D(\psi)
   \ket{\psi}_{\mathrm{mag}}\otimes\ket{K_5}_{\mathrm{mult}}.
 \label{loc:five-factorization}
\end{equation}
Here the local magnetic states $m=+1/2$ and $m=-1/2$ are identified
with logical $\ket0$ and $\ket1$, using the fixed Clebsch--Gordan
phases specified in Appendix~\ref{app:schur}. The same basis is used for
every input. The unique multiplicity factor is independent of $\psi$;
the magnetic factor contains all eight complex amplitudes of the unknown
state. This factorization follows from the multiplicity-one structure
used in Schur concentration~\cite{botero2018,gonzalez2025}, together with
$B_5=(9/10)I$ and the state-retaining map in
Theorem~\ref{alg:extra-copy}.

Taking the norm of Eq.~\eqref{loc:five-factorization} and tracing out
the multiplicity registers gives
\begin{align}
 p_5&=\frac{5t}{72},
 \label{loc:p5}\\
 \rho_{\mathrm{mag}\mid\mathrm{succ}}
 &=\ket\psi\bra\psi,\qquad t>0.
 \label{loc:retained}
\end{align}
The magnetic qubit can be transferred to a chosen output qubit by a
fixed local decoding. No estimate of $\psi$ or state-dependent correction
is needed. Measuring the magnetic labels would instead remove the
coherences that carry the unknown state.

\begin{figure*}[t]
 \includegraphics[width=\textwidth]{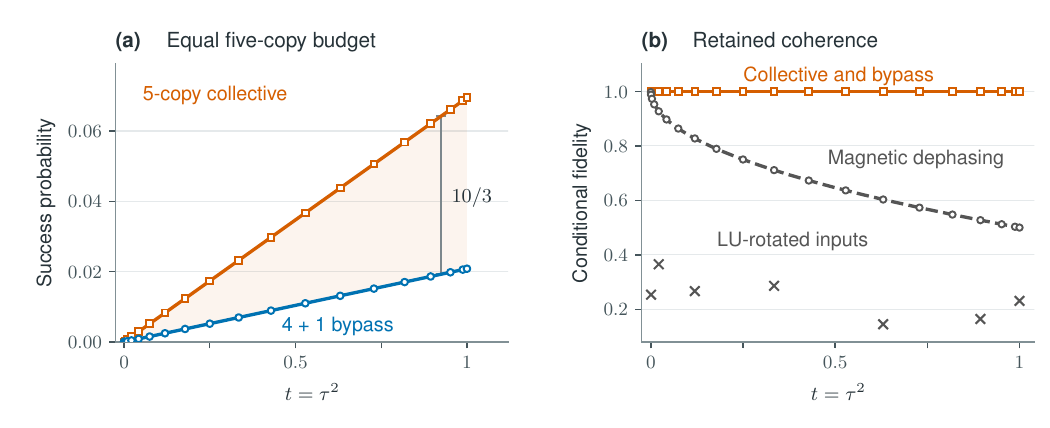}
 \caption{Five-copy gain and the quantum register that retains the input.
 (a) Both protocols consume five input copies and return $\ket\psi$ on
 success. Collective spin selection has success probability $5t/72$;
 applying the four-copy test and reserving the fifth copy gives $t/48$.
 The shaded region is the increase in success probability.
 Curves are the exact probabilities and points are physical-instrument
 calculations. (b) Conditional fidelity of the decoded magnetic register,
 with and without a nonselective measurement of its computational-basis
 labels. The states include 21 members of the complex GHZ family
 $\cos\theta\ket{000}+e^{i\pi/3}\sin\theta\ket{111}$ and seven local-unitary
 images. The coherent output has unit fidelity. Magnetic dephasing gives
 $1-\sqrt t/2$ on the GHZ family; crosses show its basis-dependent effect
 on the local-unitary images. Conditional fidelity is undefined at the
 zero-success endpoint, which is omitted from (b).}
 \label{fig:retention}
\end{figure*}

For an equal input budget of five copies, another state-retaining
protocol applies the optimal four-copy test and leaves the fifth copy
untouched. Its success probability is $t/48$. The collective five-copy
protocol increases this probability by $10/3$, while both successful
outputs have unit fidelity with the input. Figure~\ref{fig:retention}
compares these probabilities and the effect of measuring the magnetic
registers. For a nonselective computational-basis measurement of all
three retained qubits, the fidelity becomes
$\sum_{i=0}^{7}|\psi_i|^4$. In the generalized GHZ family this is
$1-\sqrt t/2$, reaching $1/2$ at maximal three-tangle. The loss of
coherence identifies the quantum register responsible for state
retention.

\subsection{Six copies and task separation}

At six copies the residual space of the multiplication map is
$\Sym^2\Hilbert$. Its local-spin decomposition has four inequivalent
irreducible sectors,
\begin{equation}
 \Sym^2\Hilbert
 =R_{111}\oplus R_{100}\oplus R_{010}\oplus R_{001},
 \label{loc:residual}
\end{equation}
of dimensions $27,3,3,3$. We also use $R_{uvw}$ for the corresponding
orthogonal projectors and set $R_s=R_{100}+R_{010}+R_{001}$.
The Gram operator has two eigenvalues,
\begin{equation}
 B_6=\frac{37}{90}R_{111}+\frac12R_s,
 \label{loc:b6}
\end{equation}
as derived in Eq.~\eqref{alg:six-spectrum}.
The weights of $\ket\psi^{\otimes2}$ in these two sectors are
\begin{equation}
 w_{111}=\frac{1+S}{4},\qquad w_s=\frac{3-S}{4}.
 \label{loc:weights}
\end{equation}
These expressions follow by expanding local swap projectors and using
$\langle F_X\rangle=\Tr\rho_X^2$; the derivation appears in
Appendix~\ref{app:schur}.

\begin{figure*}[t]
 \includegraphics[width=\textwidth]{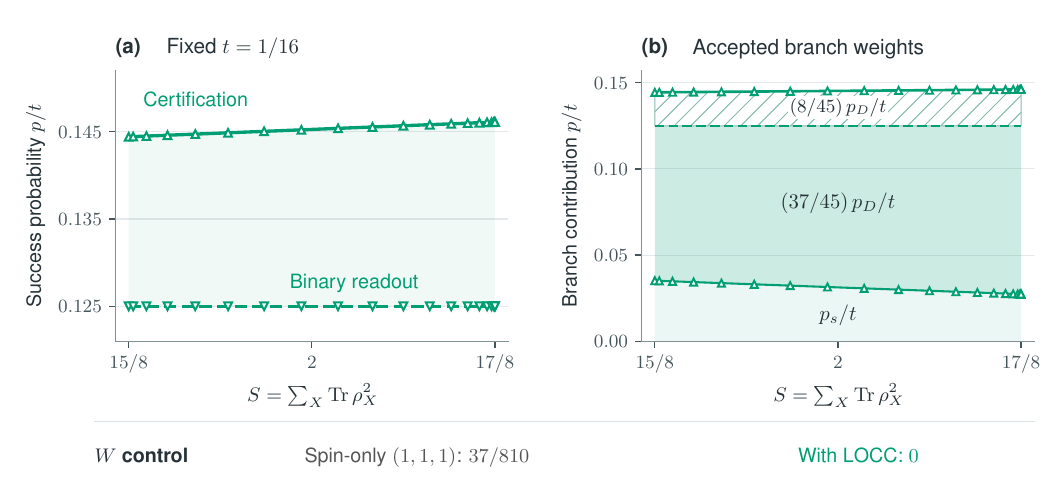}
 \caption{The first separation between optimal certification and exact binary
 readout. (a) In the fixed-$t$ family of Eq.~\eqref{loc:fixed-t-family},
 $t=1/16$ while the marginal-purity sum varies. The six-copy binary
 coefficient is constant; the certification coefficient depends on $S$.
 The shaded region is their difference.
 (b) The certification probability splits into directly accepted spin
 sectors, $p_s$, and the distinguished multiplicity branch, $p_D$.
 The hatched part, $(1-37/45)p_D$, is removed by classical thinning to
 obtain $t/8$. Lines and areas are exact expressions; points in (a) and
 (b) are instrument calculations on 21 states. The control below shows that
 accepting the entire
 $(1,1,1)$ spin sector would give the W state a false certificate with
 probability $37/810$. Conditional multiplicity discrimination makes that
 probability zero.}
 \label{fig:separation}
\end{figure*}

The optimal certification projector weights the residual sectors by
the inverse Gram eigenvalues. Its success probability is therefore
\begin{align}
 p_6^{\mathrm{cert}}
 &=\frac{t}{16}\left(\frac{90}{37}w_{111}+2w_s\right)
 =\frac{39+2S}{296}\,t,
 \label{loc:p6cert}\\
 p_6^{\mathrm{bin}}&=\frac{t}{16\lambda_{\max}(B_6)}
 =\frac{t}{8}.
 \label{loc:p6bin}
\end{align}
The first probability is optimized for each input by a single fixed
measurement; $S$ describes its response and need not be known in advance.
The second probability is constrained to encode $t$ with a
state-independent coefficient. Since $S\ge3/2$, their ratio satisfies
\begin{equation}
 \frac{p_6^{\mathrm{cert}}}{p_6^{\mathrm{bin}}}
 =\frac{39+2S}{37}\ge\frac{42}{37}>1\qquad(t>0).
 \label{loc:separation}
\end{equation}
Six is consequently the first copy number at which the two optimal
tasks differ. Figure~\ref{fig:separation} displays this distinction on
a continuous family with fixed $t=1/16$: certification varies with the
single-party purities, whereas binary readout remains constant.
The family is
\begin{align}
 \ket{\psi_\theta}={}&\frac{\ket{000}+\ket{111}}2
 +\frac{i\cos\theta}{\sqrt2}\ket{100}\nonumber\\
 &+\frac{\sin\theta}{2}(\ket{101}+\ket{110}),
 \quad 0\le\theta\le\frac\pi2,
 \label{loc:fixed-t-family}
\end{align}
with $S=15/8+\cos^2\theta/2-\cos^4\theta/4$.

\subsection{Conditional local discrimination}

The six-copy local protocol first measures $(J_A,J_B,J_C)$. It directly
accepts $(1,0,0)$ and its two permutations. Each such sector has a unique
invariant multiplicity vector. In contrast, the $(1,1,1)$ sector has
two invariant multiplicity directions~\cite{gonzalez2025}, so a
Schur-decoded input in this sector takes the form
\begin{equation}
 \ket{x_D(\psi)}_{\mathrm{mag}}\ket{K_D}_{\mathrm{mult}}
 +\ket{x_0(\psi)}_{\mathrm{mag}}\ket{K_0}_{\mathrm{mult}}.
 \label{loc:six-factorization}
\end{equation}
The fixed normalized vectors $K_D,K_0\in(\mathbb C^9)^{\otimes3}$
are orthogonal; only $K_D$ belongs to the certification subspace.
Accepting the entire spin sector would accept the W state with
probability $37/810$ despite $D(W)=0$.

The two multiplicity states can be distinguished perfectly by
LOCC~\cite{walgate2000}. Let $E_D$ be the effect associated with the
outcome $K_D$. Perfect discrimination gives
$\bra{K_D}E_D\ket{K_D}=1$ and
$\bra{K_0}E_D\ket{K_0}=0$. Positivity of both $E_D$ and $I-E_D$
then implies
\begin{equation}
 E_D\ket{K_D}=\ket{K_D},\qquad E_D\ket{K_0}=0.
 \label{loc:full-effect}
\end{equation}
Thus $I_{\mathrm{mag}}\otimes E_D$ accepts exactly the first
component of Eq.~\eqref{loc:six-factorization}, including when the two
registers are correlated. Combining this measurement with the three
directly accepted sectors implements $P_6$ on the whole allowed input
space. Appendix~\ref{app:locc} specifies fixed local bases and the
Alice-to-Bob-to-Charlie communication that realize the effect.

The two types of successful branches have probabilities
\begin{equation}
 p_D=\frac{45(1+S)}{1184}\,t,\qquad
 p_s=\frac{3-S}{32}\,t.
 \label{loc:branch-probabilities}
\end{equation}
Certification accepts both. To obtain the optimal binary readout, one
accepts the $K_D$ branch with probability $a=37/45$ and always accepts
the other three successful sectors. The resulting probability is
\begin{equation}
 a p_D+p_s=\frac{t}{8}.
 \label{loc:thinning}
\end{equation}
The quantum operations are shared by the two tasks; they differ only
in this final classical acceptance rule. Retaining the branch record
also permits a weighted estimator without the extra randomization,
as discussed below.

\section{Copy cost and statistical readout}
\label{sec:statistics}

\subsection{Certification at a fixed input budget}

For independent batches with success probability $p>0$, the number of batches
to the first success is geometric. The expected number of input states is
\begin{equation}
 \mathbb E[C_{\rm first}]=\frac{n}{p}.
 \label{eq:first-success}
\end{equation}
The four-copy and five-copy protocols therefore require $192/t$ and $72/t$
input copies, respectively. Their input-cost ratio is $8/3$. For the
four-copy measurement with a fifth copy reserved, the cost is instead $240/t$;
the collective five-copy protocol reduces this cost by $10/3$ while producing
the same retained output. Six-copy binary readout costs $48/t$, and six-copy
certification costs $1776/[(39+2S)t]$.

A predetermined input budget $C$ permits $N=\lfloor C/n\rfloor$ batches of a
fixed protocol. Its probability of returning no certificate is $(1-p)^N$.
For a guaranteed alternative $\tau\ge\tau_0>0$, a binary protocol with
coefficient $\kappa$ reaches miss probability at most $\delta$ using
\begin{equation}
 N\ge\left\lceil\frac{\log\delta}
 {\log(1-\kappa\tau_0^2)}\right\rceil.
 \label{eq:miss-budget}
\end{equation}
Six-copy certification can use the uniform lower coefficient $21/148$,
since $S\ge3/2$. Figure~\ref{fig:statistics} compares the exact miss
probabilities for a maximally entangled GHZ state and a weakly entangled
member of the same family. The state-dependent coefficient is used when
reporting a particular input's performance.

\subsection{Binary and weighted readout}

For $X\sim\operatorname{Binomial}(N,\kappa t)$, the estimator
\begin{equation}
 \widehat t=\frac{X}{N\kappa},\qquad
 \operatorname{Var}(\widehat t)=\frac{t(1-\kappa t)}{N\kappa}
 \label{eq:binary-estimate}
\end{equation}
is unbiased. Clipping it to $[0,1]$ changes its bias, as does taking a square
root to estimate $\tau$. All variance comparisons here concern the raw
estimator of $t$.

For six copies, let $X_D$ count the accepted $(1,1,1)$ multiplicity direction
and $X_s$ count the three $(1,0,0)$-type branches. Write their probabilities
as $p_D,p_s$, and set $a=37/45$. The entire branch record gives
\begin{align}
 \widehat t_{\rm w}&=\frac8N(aX_D+X_s),\label{eq:weighted-estimate}\\
 \operatorname{Var}(\widehat t_{\rm w})
 &=\frac{64}{N}\left[a^2p_D+p_s-(ap_D+p_s)^2\right].
 \label{eq:weighted-variance}
\end{align}
This estimator is unbiased because $ap_D+p_s=t/8$. Conditional on the
record, replacing an accepted $D$ outcome by an independent Bernoulli trial
with mean $a$ adds variance. The difference between binary thinning and
weighted readout is exactly
\begin{equation}
 \operatorname{Var}(\widehat t_{\rm bin})-
 \operatorname{Var}(\widehat t_{\rm w})
 =\frac{64}{N}a(1-a)p_D.
 \label{eq:variance-gap}
\end{equation}
The weighted estimator is the conditional expectation of the randomized
binary estimator. Thus the task separation in Eq.~\eqref{eq:task-gap}
coexists with a useful quantitative estimate from every recorded success
branch.

\begin{figure*}[t]
 \includegraphics[width=\textwidth]{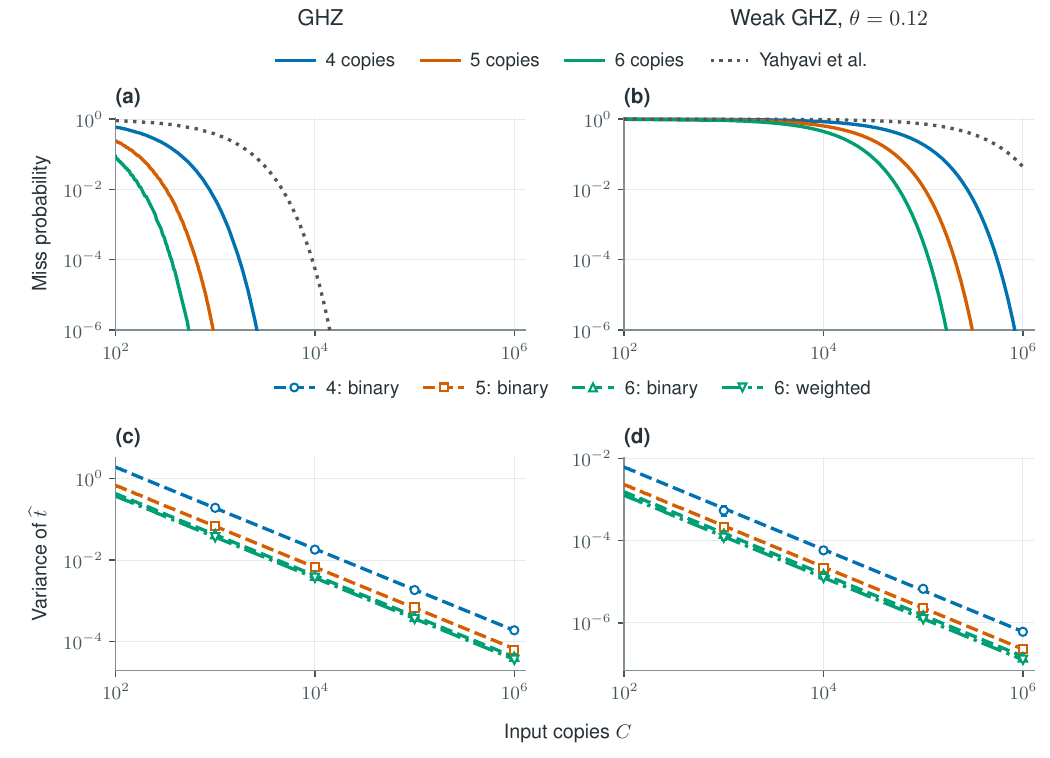}
 \caption{Input-copy budgets for certification and estimation. Columns use
 the GHZ state ($t=1$) and the weak GHZ state
 $\cos(0.12)\ket{000}+e^{0.37i}\sin(0.12)\ket{111}$
 ($t\simeq0.00319$). (a),(b) Exact probability of no certificate after
 $N=\lfloor C/n\rfloor$ independent batches. The shallow four-copy reference
 has success probability $t/256$~\cite{yahyavi2022}.
 (c),(d) Variance of the raw unbiased estimator of $t$ under the same input
 budget. Dashed curves use binary outcomes; the six-copy dash-dotted curve
 uses the full weighted branch record. Points are sample variances from
 1000 repetitions, with error bars of one Monte Carlo standard error of
 the sample variance. They describe uncertainty in the variance estimate.
 Weighting improves on binary thinning without increasing the quantum
 measurement cost.}
 \label{fig:statistics}
\end{figure*}

The variance curves in Fig.~\ref{fig:statistics} use
Eqs.~\eqref{eq:binary-estimate} and \eqref{eq:weighted-variance}. Points are
sample variances from 1000 independent multinomial or binomial simulations
per configuration. Equal-tailed binomial intervals and a simultaneous
interval for the weighted statistic are given in
Appendix~\ref{app:statistics}; the latter uses the two integer branch counts.

\subsection{Comparison with reconstruction methods}

Table~\ref{tab:methods} separates simultaneous block size from the statistic
that is read out. For a learned measurement basis, the total budget is
\begin{equation}
 C=C_{\rm train}+C_{\rm cal}+nN+C_{\rm unused},
 \label{eq:total-budget}
\end{equation}
where training includes unsuccessful optimization attempts. A fixed local
spin filter uses neither a learned basis nor state-dependent control gates.

We also compare with local Pauli tomography and the canonical-form method
of P\'erez-Salinas \emph{et al.}~\cite{perezsalinas2020}. Tomography allocates
copies among the 27 local Pauli settings, then projects a linear-inversion
estimate onto its leading pure-state eigenvector. The canonical-form
implementation allocates 70\% of the budget to a sampled optimization and
uses the remaining copies for readout. Its final statistic estimates the
tangle to the accuracy with which the canonical form has been reached.

For one complex input with $t=0.49931$ and $S=1.68521$, at $C=10^4$ the
observed mean squared errors of the four-copy, five-copy, and six-copy
weighted estimators are $9.82\times10^{-3}$, $3.47\times10^{-3}$, and
$1.98\times10^{-3}$. The pure-state tomography estimate gives
$4.84\times10^{-4}$. These values use 1000 repetitions for the event
estimators and 30 for tomography. The comparison shows that optimizing a
conclusive event and reconstructing a numerical invariant favor different
measurement statistics. In the same sampled canonical-form experiment,
none of the 30 final readouts established a residual objective below
$0.01$ at 95\% confidence; its errors include the remaining basis-learning error
(Appendix~\ref{app:statistics}).

\begin{table*}[t]
 \caption{Readout tasks and the required coherent block size. ``Learned''
 denotes a measurement basis adjusted using the unknown input. A retained
 state is an explicit output of the indicated protocol. The four-copy
 singlet product and its shallow circuit give the same probability coefficient;
 selecting the complete local singlet sector gives the stronger reference
 probability $t/48$.}
 \label{tab:methods}
 \begin{ruledtabular}
 \begin{tabular}{l c c l l}
 Method & Copies per block & Basis & Readout statistic & Retained state\\
 \hline
 Singlet product~\cite{yu2007}; circuit~\cite{yahyavi2022}
 &4&Fixed&Success probability $t/256$&---\\
 Complete singlet sector~\cite{sugita2008}
 &4&Fixed&Success probability $t/48$&---\\
 Basis-adjusted reconstruction~\cite{yu2016}
 &2&Learned&Several probabilities&---\\
 Canonical-form readout~\cite{perezsalinas2020}
 &1&Learned&$\tau=4P_{000}P_{111}$ in canonical form&---\\
 Ancilla readout~\cite{liu2026}
 &4&Fixed&Reported success probability $t/2048$&---\\
 Five-copy local protocol&5&Fixed&Success probability $5t/72$&$\ket\psi$\\
 Six-copy local protocol&6&Fixed&Binary $t/8$ or weighted record&---
 \end{tabular}
 \end{ruledtabular}
\end{table*}

\section{Circuit resources and robustness}
\label{sec:resources}

\subsection{Local circuit resources}

The local protocols require coherent processing only within a party.
Schur and Clebsch--Gordan transforms provide systematic circuit constructions
for this operation~\cite{bacon2006,kirby2018}. We compile explicit local
unitaries into arbitrary single-qubit rotations and CX gates, including
ancillary spin flags and all conditional discrimination gates. Shared binary
conditions within each accepted spin sector are evaluated by coherent
controlled flips. We consider
all-to-all connectivity and a nearest-neighbor line within each party.
No gate connects different parties.

\begin{figure*}[t]
 \includegraphics[width=\textwidth]{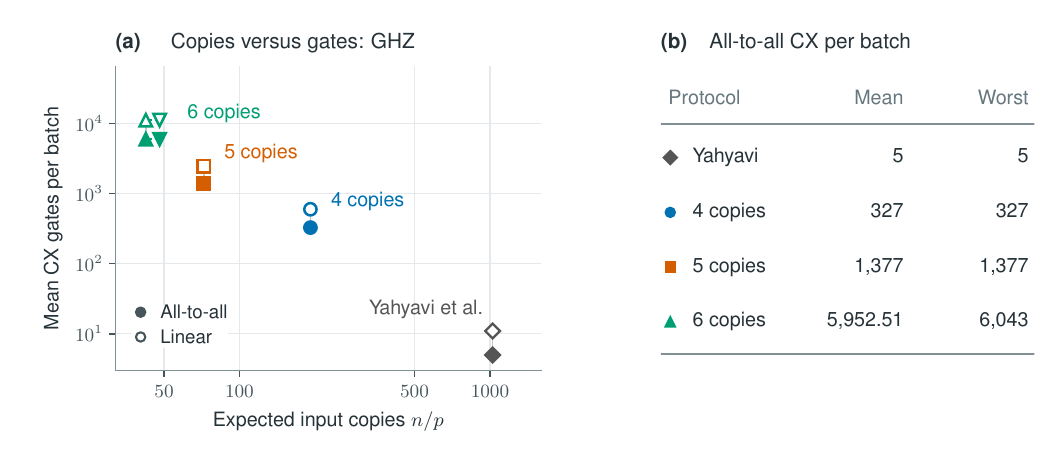}
 \caption{Copy savings and coherent gate costs for a GHZ input.
 (a) Expected input copies to the first success, $n/p$, on the horizontal
 axis and mean CX gates per complete batch on the vertical axis.
 Filled and open markers denote
 all-to-all and linear connectivity within each party; thin gray lines
 connect the two implementations. Upward and downward six-copy triangles
 denote certification and binary readout, respectively. Six-copy
 certification and binary readout share quantum operations but have
 different acceptance probabilities. The shallow four-copy circuit is
 from Ref.~\cite{yahyavi2022}. (b) The aligned columns give average and
 worst-path CX counts of the all-to-all implementations.
 The six-copy average includes failed and
 successful batches; the difference from the worst path reflects
 conditional execution, not statistical error. Counts include all parties
 and auxiliary flag operations but exclude state preparation.}
 \label{fig:resources}
\end{figure*}

For all-to-all connectivity, the complete four-copy and five-copy filters
use 327 and 1377 CX gates, respectively. The six-copy protocol uses 5952.51
CX gates on average for a GHZ input and at most 6043 along an execution
path. The average includes both successful and rejected outcomes. The
corresponding shallow four-copy circuit of Yahyavi
\emph{et al.}~\cite{yahyavi2022} uses five CX gates in total, with success
probability $t/256$. Figure~\ref{fig:resources} places this gate cost beside
the expected input cost. The collective protocols reduce the number of
preparations needed for a certificate, at the price of more coherent gates
in these implementations.

Six-copy branches execute different conditional operations. If $G$ is
the number of gates used by a batch, including a batch that fails, repeated
independent execution has expected gate cost $\mathbb E[G]/p$ to the first
success. Using only the cost of successful branches would omit the dominant
expense when $p$ is small. Circuit depths allow the three initial local
filters to run in parallel and then order the conditional Alice--Bob--Charlie
measurements. Classical communication latency is a separate physical cost.
Appendix~\ref{app:implementation} specifies the registers and resource
conventions.

\subsection{Preparation errors and conditional output}

The purity and identical-copy assumptions have a direct operational role.
Let the actual batch be $\rho_1\otimes\cdots\otimes\rho_n$, and suppose
$T(\rho_r,\ket\phi\bra\phi)\le\eta_r$ for the same null state $\phi$, where
$T(\rho,\sigma)=\|\rho-\sigma\|_1/2$. If the implemented effect differs from
the ideal full-space effect by at most $\epsilon_M$ in operator norm, then
\begin{equation}
 p_{\rm false}\le\min\left\{1,\sum_{r=1}^n\eta_r+\epsilon_M\right\}.
 \label{eq:robust-bound}
\end{equation}
For an identical coherently perturbed pure state with trace distance at most
$\eta\in[0,1]$ from $\phi$, the ideal-effect contribution improves to
$1-(1-\eta^2)^n\le n\eta^2$. The proofs and channel definitions are given in
Appendix~\ref{app:implementation}.

The physical extension of an effect matters under mixed preparation noise.
For example, the five-copy minimal-support global projector accepts five
maximally mixed three-qubit states with probability $1/4096$. The local
spin-$1/2$ filter accepts the same input with probability $125/4096$, since
each local accepted sector has rank ten. Both effects give $5t/72$ on
identical pure inputs. Noise calculations must therefore act on the
specified physical instrument.

Conditional state quality also depends on the success map. Let its ideal
unnormalized output be $\sigma=p\ket\psi\bra\psi$ and its actual output be
$\widetilde\sigma$, with
$\|\widetilde\sigma-\sigma\|_1\le\epsilon_s$ and
$\widetilde p=\Tr\widetilde\sigma>0$. Then
\begin{equation}
 T\!\left(\frac{\widetilde\sigma}{\widetilde p},
       \ket\psi\bra\psi\right)
 \le \min\{1,\epsilon_s/\widetilde p\}.
 \label{eq:retention-robustness}
\end{equation}
Postselection can thus amplify errors at low success probability. Acceptance
probability and retained-state fidelity are distinct observables.

With a calibrated null bound $p_0$, a fixed-batch experiment can instead
certify at a controlled error rate through the upper binomial tail
$\Pr_{p_0}(X\ge x)$. This connects the ideal conclusive event to a noisy
statistical test. The null model, calibration cost, and test error must
be specified together, as detailed in Appendix~\ref{app:statistics}.

\section{Conclusions}
\label{sec:conclusions}

The copy cost of three-qubit entanglement readout depends on the required
output. An algebraic null set fixes the optimal conclusive support, while
the spectrum of polynomial multiplication fixes the largest exact binary
probability. These two descriptions coincide at four and five copies.
The fifth copy increases the success probability by $10/3$ and carries a
recoverable copy of the unknown input through the measurement. At six
copies, the two Gram eigenvalues make certification sensitive to local
purities as well as to the three-tangle. Conditional local discrimination
attains the certification projector, and a classical acceptance rule gives
the optimal binary event.

Keeping the complete success record provides a weighted estimator without
the extra variance of binary thinning. Comparing this information with
retained quantum output and compiled gate counts gives a concrete account
of what additional copies provide. Further questions include the higher-copy
Gram spectrum, optimal retention of several unknown copies, and local
encodings that reduce the coherent cost of the optimal effects.

\begin{acknowledgments}
This work is being supported by the following: the National Natural Science Foundation of China (NSFC) under Grants 11761073, 12075159, and 12171044; the Academician Innovation Platform of Hainan Province. The authors acknowledge the assistance of generative-AI tools for gure preparation and language
re nement.
\end{acknowledgments}

\appendix
\section{Algebraic and spectral proofs}
\label{app:algebra}

\subsection{The null ideal and the four-copy threshold}

Write the amplitudes of a three-qubit vector in binary order as
$(a,b,c,d,e,f,g,h)$. Cayley's hyperdeterminant is
\begin{align}
 D={}&a^2h^2+b^2g^2+c^2f^2+d^2e^2\nonumber\\
 &-2(ahbg+ahcf+ahde+bgcf+bgde+cfde)\nonumber\\
 &+4(adfg+hbce).
 \label{alg:hyperdeterminant}
\end{align}
Consider it as a polynomial in $h$ over the unique factorization domain
$R=\mathbb C[a,b,c,d,e,f,g]$:
\begin{align}
 D&=a^2h^2+Lh+C,\nonumber\\
 L&=-2a(bg+cf+de)+4bce,\nonumber\\
 C&=b^2g^2+c^2f^2+d^2e^2\nonumber\\
  &\hspace{1em}-2bcfg-2bdeg-2cdef+4adfg.
 \label{alg:quadratic}
\end{align}
Its discriminant factors as
\begin{equation}
 L^2-4a^2C=-16(ad-bc)(af-be)(ag-ce).
 \label{alg:discriminant}
\end{equation}
Each factor is irreducible. For example, $ad-bc$ is primitive as a
polynomial in $d$, since $a$ and $bc$ are coprime, and it is linear over
the fraction field of the other variables. Gauss's lemma proves its
irreducibility in $R$; the other two factors follow by the same
argument. Their different variable dependence makes them pairwise
nonassociate. Equation~\eqref{alg:discriminant} thus has odd valuation
at each of three irreducible factors and is not a square in the
fraction field of $R$.

A quadratic over a field of characteristic zero is reducible if and
only if its discriminant is a square. Hence $D$ is irreducible over
that fraction field. It is also primitive over $R$: any common
irreducible divisor of its coefficients would divide $a^2$, whereas
$a$ does not divide $L$. A second use of Gauss's lemma proves
irreducibility in $\mathbb C[a,b,c,d,e,f,g,h]$. Its principal ideal is
therefore prime, and the Hilbert Nullstellensatz gives
\begin{equation}
 I(\{D=0\})=\sqrt{(D)}=(D).
 \label{alg:principal-ideal}
\end{equation}

For completeness, the occupation expansion
\eqref{alg:occupation} makes the map
$v\mapsto\bra{v}\ket{z}^{\otimes n}$ an antilinear bijection from
$\Hilbert_n$ to the degree-$n$ homogeneous polynomials. A vector is
orthogonal to every null tensor power precisely when its polynomial
vanishes on the null cone. This proves the ideal correspondence in
Theorem~\ref{alg:certification}, including for cones whose ideal has
several generators. In the present case,
\begin{equation}
 I(\{D=0\})_n=
 \begin{cases}
  \{0\},&n<4,\\
  D\,\mathbb C[z_0,\ldots,z_7]_{n-4},&n\ge4.
 \end{cases}
 \label{alg:ideal-degree}
\end{equation}
Multiplication by $D$ is injective, so the dimension of the second
line is $\binom{n+3}{7}$. This proves Eq.~\eqref{alg:ghz-rank}.
The use of the vanishing ideal is essential: replacing an equation
$f=0$ by $f^2=0$ leaves this ideal, and hence the threshold, unchanged.

\subsection{Uniqueness of a compressed probability effect}

Let $A$ be Hermitian on $\Hilbert_n$ and suppose
$\bra{\psi}^{\otimes n}A\ket{\psi}^{\otimes n}=0$ for every normalized
$\psi$. Homogeneity extends this identity to every $z\in\mathbb C^d$.
In the occupation basis it reads
\begin{equation}
 \sum_{|\alpha|=|\beta|=n}
 \sqrt{\binom n\alpha\binom n\beta}\,
 A_{\alpha\beta}\bar z^\alpha z^\beta=0.
 \label{alg:polarization}
\end{equation}
Set $z=x+iy$ with $x,y\in\mathbb R^d$. A polynomial vanishing for all
real $x,y$ has all coefficients zero. The complex linear substitution
$(x,y)\mapsto(z,\bar z)$ is invertible, so the monomials
$\bar z^\alpha z^\beta$ in Eq.~\eqref{alg:polarization} are linearly
independent as functions on $\mathbb C^d$. Every matrix entry of $A$
therefore vanishes.

In particular, the equality
$p_M(\psi)=c_n|f(\psi)|^2$ and
Eq.~\eqref{alg:contraction} force
$\Pi_nM\Pi_n=c_nV_nV_n^\dagger$.
The nonzero eigenvalues of $V_nV_n^\dagger$ are those of $B_n$.
The contraction condition gives
$c_n\lambda_{\max}(B_n)\le1$; extending the effect by zero on
$\Hilbert_n^\perp$ attains equality. No ideal assumption is needed
for this binary-readout result.

When $I(Z)=(f)$, the polynomial associated with $V_nw$ is $f$ times
the polynomial associated with $w$, so
$\operatorname{ran}V_n=\mathcal K_n^\perp$.
The full column rank of $V_n$ gives the projector
$V_nB_n^{-1}V_n^\dagger$, establishing
Eqs.~\eqref{alg:gram-projector} and \eqref{alg:cert-probability}.
If $V_n=U\Sigma W^\dagger$ is a thin singular-value decomposition, the
certification and optimal binary effects on this range are,
respectively, $UU^\dagger$ and
$U\Sigma^2U^\dagger/\lambda_{\max}(B_n)$.
They coincide if and only if $B_n$ is scalar.

\subsection{The additional-copy constant and its instrument}

Let $\ket{v_f}=V_m1$, so
\begin{equation}
 \ket{v_f}=\sum_{|\alpha|=m}
 \frac{\overline{c_\alpha}}{\sqrt{\binom m\alpha}}
 \ket{\alpha},\qquad \|v_f\|^2=q_f.
 \label{alg:coefficient-vector}
\end{equation}
Denote the action of $U$ on $\Sym^r\Hilbert$ by $U_r$. Applying
Eq.~\eqref{alg:contraction} to $Uz$ proves
\begin{equation}
 U_nV_n=\chi(U)V_nU_{n-m}.
 \label{alg:intertwiner}
\end{equation}
Pure powers span each symmetric tensor space, so this equality holds
as an operator identity. Since $|\chi(U)|=1$, it implies that $B_n$
commutes with $U_{n-m}$. At $n=m+1$, irreducibility of the original
representation and Schur's lemma give $B_{m+1}=bI$.

The scalar follows from a trace that retains the occupation
normalization explicitly. Let $a_j^\dagger$ be bosonic creation
operators for the $d$ basis modes. Comparing coefficients in
Eq.~\eqref{alg:multiplication} gives
\begin{equation}
 V_{m+1}\ket{j}
 =\frac{a_j^\dagger\ket{v_f}}{\sqrt{m+1}}.
 \label{alg:creation-one}
\end{equation}
On the $m$-particle sector,
$\sum_j a_ja_j^\dagger=(m+d)I$. Thus
\begin{equation}
 db=\Tr B_{m+1}
 =\frac{1}{m+1}\sum_j\|a_j^\dagger v_f\|^2
 =\frac{m+d}{m+1}q_f.
 \label{alg:gram-trace}
\end{equation}
Together with $B_m=q_f$, this proves the coefficient and gain in
Theorem~\ref{alg:extra-copy}.

Extend $K=V_{m+1}^\dagger/\sqrt b$ by zero outside
$\Hilbert_{m+1}$. Since $B_{m+1}=bI$,
$K^\dagger K=V_{m+1}V_{m+1}^\dagger/b$ is a projector on its input
space. It can be completed to a quantum instrument by a failure
operator with effect $I-K^\dagger K$. The success vector is precisely
Eq.~\eqref{alg:retention-kraus}. On every input with $f(\psi)\ne0$,
the conditional output density operator is
$\ket{\psi}\bra{\psi}$. This construction is global; locality of a
particular implementation is a separate property established for $D$
in Sec.~\ref{sec:local}.

To prove Corollary~\ref{alg:retention-minimum}, let
$\ket{u}=\ket{v_D}/\sqrt{q_D}$ span the four-copy certification
space. For any successful instrument with Kraus operators $K_\mu$,
the effect $\sum_\mu K_\mu^\dagger K_\mu$ annihilates
$\mathcal K_4$. Positivity implies that every $K_\mu$ annihilates it
separately. Consequently
\begin{equation}
 K_\mu\Pi_4=\ket{w_\mu}\bra{u},\qquad
 K_\mu\ket{\psi}^{\otimes4}
 =\frac{D(\psi)}{\sqrt{q_D}}\ket{w_\mu}.
 \label{alg:four-copy-output}
\end{equation}
After conditioning on success, the output is proportional to
$\sum_\mu\ket{w_\mu}\bra{w_\mu}$ and independent of $\psi$.
Appending a classical outcome register only replaces $w_\mu$ by a
vector with that register, leaving the conclusion unchanged.
Therefore neither feedback nor retaining the outcome can produce
the varying input state on this success branch. The five-copy
instrument above supplies sufficiency.

\subsection{Exact Gram spectra through six copies}

For $D$, the four square terms, six terms of coefficient $-2$, and
two terms of coefficient $4$ in Eq.~\eqref{alg:hyperdeterminant} give
\begin{equation}
 q_D=\frac{4}{6}+\frac{6\cdot4}{24}
                  +\frac{2\cdot16}{24}=3.
 \label{alg:d-norm}
\end{equation}
The product $\mathrm{SU}(2)^3$ representation on $\mathbb C^8$ is
irreducible. Its invariance of $D$ follows directly from the two matrix
slices $A_0=\left(\begin{smallmatrix}a&b\\c&d\end{smallmatrix}\right)$
and $A_1=\left(\begin{smallmatrix}e&f\\g&h\end{smallmatrix}\right)$:
$D$ is the discriminant of $\det(xA_0+yA_1)$. A determinant-one
transformation at $B$ or $C$ leaves this binary quadratic unchanged;
a determinant-one transformation at $A$ changes its variables and
preserves its discriminant. Equations
\eqref{alg:extra-gram} and \eqref{alg:d-norm} yield $B_4=3$ and
$B_5=(9/10)I_8$.

At six copies the residual space is $\Sym^2\Hilbert$. A local pair
of qubits is either a spin-one triplet, symmetric under exchange, or
a spin-zero singlet, antisymmetric under exchange. Overall symmetry
requires an even number of local singlets. Hence the residual space
decomposes without multiplicities as
\begin{equation}
 \Sym^2\Hilbert
 =R_{111}\oplus R_{100}\oplus R_{010}\oplus R_{001},
 \label{alg:residual-decomposition}
\end{equation}
where subscripts specify the three local spins; the dimensions are
$27$, $3$, $3$, and $3$. We also use $R_{111}$ for its projector and
write $R_s=R_{100}+R_{010}+R_{001}$ for the complementary projector.
Covariance makes $B_6$ scalar on each irreducible summand. Permutation
symmetry among the parties makes its scalar the same on the last
three summands. It remains to evaluate two matrix elements.

Label the eight bosonic modes by $0,\ldots,7$ in binary order and set
$v=v_D$. Comparing occupation coefficients as in
Eq.~\eqref{alg:creation-one} gives
\begin{align}
 V_6\ket{2_i}&=\frac{(a_i^\dagger)^2v}{\sqrt{30}},\nonumber\\
 V_6\ket{1_i1_j}&=\frac{a_i^\dagger a_j^\dagger v}{\sqrt{15}}
 \quad(i\ne j).
 \label{alg:creation-two}
\end{align}
The vector $\ket{2_0}=\ket{000}^{\otimes2}$ belongs to $R_{111}$.
The squared coefficients of $v$ with occupation $\alpha_0=0,1,2$
sum to $5/3$, $7/6$, and $1/6$, respectively. Therefore
\begin{equation}
 \|(a_0^\dagger)^2v\|^2
 =2\frac53+6\frac76+12\frac16=\frac{37}{3},
 \label{alg:triplet-norm}
\end{equation}
giving the eigenvalue $37/90$ on $R_{111}$.

A normalized vector in $R_{100}$ is
\begin{equation}
 \ket{u_s}=
 \frac{\ket{1_01_3}-\ket{1_11_2}}{\sqrt2}.
 \label{alg:singlet-test-vector}
\end{equation}
In the ordering by parties, it is a spin-one highest-weight vector
at $A$ and a singlet at each of $B$ and $C$. Equation
\eqref{alg:creation-two} gives
$V_6u_s=(a_0^\dagger a_3^\dagger-a_1^\dagger a_2^\dagger)v/\sqrt{30}$.
The coefficients in Eq.~\eqref{alg:hyperdeterminant} yield
\begin{align}
 \|a_0^\dagger a_3^\dagger v\|^2
 &=\|a_1^\dagger a_2^\dagger v\|^2=\frac{41}{6},\nonumber\\
 \langle a_0^\dagger a_3^\dagger v,
         a_1^\dagger a_2^\dagger v\rangle&=-\frac23.
 \label{alg:singlet-norms}
\end{align}
These contractions can be evaluated directly by assigning to a
monomial of exponent $\gamma$ in
$z_i z_jD(z)$ the amplitude
$(\prod_k\gamma_k!)^{1/2}/\sqrt{4!}$ times its coefficient.
Taking squared norms or matching equal exponents gives
Eq.~\eqref{alg:singlet-norms} without diagonalizing a matrix.
It follows that $\|V_6u_s\|^2=(41/3+4/3)/30=1/2$.
Combining the two sectors proves
\begin{equation}
 \boxed{B_6=\frac{37}{90}R_{111}+\frac12 R_s,}
 \qquad
 \begin{matrix}
  \rank R_{111}=27,\\
  \rank R_s=9.
 \end{matrix}
 \label{alg:six-spectrum}
\end{equation}
In particular, the optimal binary coefficient is $1/8$.
The two distinct eigenvalues also establish that the optimal
six-copy certification projector differs from the optimal binary
effect. The sector weights of an identical residual pair, and the
local measurements resolving the corresponding success branches,
are derived in Appendix~\ref{app:schur}.

\section{Schur factorization and the retained state}
\label{app:schur}

\subsection{Basis and phase conventions}

We use computational $\ket0$ for $m=+1/2$ and $\ket1$ for $m=-1/2$.
A coupling path $\alpha=(j_1,\ldots,j_n)$ starts at $j_1=1/2$ and
obeys $j_{r+1}=j_r\pm1/2\ge0$. The following real Clebsch--Gordan
recursion fixes every column of the local Schur embedding:
\begin{align}
 \ket{j+\tfrac12,M;\alpha,+}
 &=\sqrt{\frac{j+M+1/2}{2j+1}}
   \ket{j,M-\tfrac12;\alpha}\ket0\nonumber\\
 &\quad+\sqrt{\frac{j-M+1/2}{2j+1}}
   \ket{j,M+\tfrac12;\alpha}\ket1,
 \label{loc:cg-plus}\\
 \ket{j-\tfrac12,M;\alpha,-}
 &=-\sqrt{\frac{j-M+1/2}{2j+1}}
   \ket{j,M-\tfrac12;\alpha}\ket0\nonumber\\
 &\quad+\sqrt{\frac{j+M+1/2}{2j+1}}
   \ket{j,M+\tfrac12;\alpha}\ket1.
 \label{loc:cg-minus}
\end{align}
A term with an unavailable magnetic label is zero. For a fixed final
spin $J$, let $E_J^{(n)}$ have columns $\ket{J,m;\alpha}$ ordered by
decreasing $m$ and then lexicographic coupling path. It is an isometry,
and $Q_J^{(n)}=E_J^{(n)}E_J^{(n)\dagger}$. All three parties use this
same convention and retain the original ordering of their local
copies.

For a triple of spins $\boldsymbol J$, define the sector decoder
\begin{equation}
 \mathcal S_{\boldsymbol J}^{(n)}
 = E_{J_A}^{(n)\dagger}\otimes
   E_{J_B}^{(n)\dagger}\otimes E_{J_C}^{(n)\dagger},
 \label{loc:sector-decoder}
\end{equation}
followed by a fixed reordering that places the three magnetic registers
before the three multiplicity registers. It maps the physical tensor
ordered by party to this decoded sector and vanishes on its orthogonal
complement.

\subsection{Invariant multiplicities}

The subspace $\Sym^n\Hilbert$ is invariant under synchronous
permutations of the $n$ copies held by all parties. In a fixed local
spin sector this condition acts only on the multiplicity registers.
Its dimension there is the Kronecker coefficient
\begin{equation}
 g(\lambda_A,\lambda_B,\lambda_C)
 =\sum_{\mu\vdash n}
   \frac{\chi_{\lambda_A}(\mu)\chi_{\lambda_B}(\mu)
                         \chi_{\lambda_C}(\mu)}{z_\mu},
 \label{loc:character-average}
\end{equation}
where $z_\mu=\prod_\ell \ell^{m_\ell}m_\ell!$ for a cycle type with
$m_\ell$ cycles of length $\ell$. This is the trace of the permutation
group average.

The characters needed here can be computed without constructing the
permutation matrices. For a permutation with cycle lengths
$\ell_1,\ldots,\ell_r$, let $F_\mu(x)=\prod_a(1+x^{\ell_a})$.
The coefficient of $x^k$ counts fixed computational states with $k$
down spins. Subtracting adjacent magnetic-weight spaces isolates a
single spin multiplicity representation, giving
\begin{equation}
 \chi_{(n-k,k)}(\mu)
 =[x^k]F_\mu(x)-[x^{k-1}]F_\mu(x).
 \label{loc:characters}
\end{equation}
For four and five copies, the nonzero contributions to
Eq.~\eqref{loc:character-average} are respectively
\begin{align}
 g((2,2)^3)&=-\frac13+1+\frac13=1,\nonumber\\
 g((3,2)^3)&=-\frac14+\frac16-\frac16+\frac18
              +\frac1{12}+\frac{25}{24}=1.
 \label{loc:g45}
\end{align}
For six copies they give
\begin{align}
 g((4,2)^3)
 &=-\frac15+\frac18-\frac18+\frac9{16}\nonumber\\
 &\quad+\frac1{16}+\frac9{16}+\frac{81}{80}=2,\nonumber\\
 g((4,2),(3,3),(3,3))
 &=\frac18-\frac18+\frac9{16}\nonumber\\
 &\quad+\frac1{16}+\frac1{16}+\frac5{16}=1.
 \label{loc:g6}
\end{align}
Here $(\lambda)^3$ denotes three copies of the partition $\lambda$.
The local multiplicity dimensions, obtained from the identity
permutation in Eq.~\eqref{loc:characters}, are $2$ for $(2,2)$, $5$
for $(3,2)$, $9$ for $(4,2)$, and $5$ for $(3,3)$. The one- and
two-dimensional invariant spaces agree with the Kronecker-state
construction of Ref.~\cite{gonzalez2025}.

\subsection{Identification of the optimal local sectors}

For $G=SU(2)_A\times SU(2)_B\times SU(2)_C$, invariance of $D$ and
the defining relation $V_n^\dagger\ket\psi^{\otimes n}
=D(\psi)\ket\psi^{\otimes(n-4)}$ imply the intertwining relation
\begin{equation}
 U^{\otimes n}V_n=V_n U^{\otimes(n-4)}
 \quad\text{on }\Sym^{n-4}\Hilbert,
 \label{loc:intertwiner}
\end{equation}
with $U=U_A\otimes U_B\otimes U_C$. The operator $V_n$ is injective.
For $n=4$, its image is a one-dimensional $G$ singlet. Equation
\eqref{loc:g45} shows that the three local spin-zero sectors have
exactly this one-dimensional intersection with $\Sym^4\Hilbert$.
Thus the compressed local projector equals $P_4$.

For $n=5$, the domain of $V_5$ is the irreducible representation
$(1/2,1/2,1/2)$ of dimension eight. Its image must lie in that same
local spin type. Equation~\eqref{loc:g45} shows that this type occurs
once in $\Sym^5\Hilbert$, with decoded form
\begin{equation}
 \mathcal U_{1/2}^{\otimes3}\otimes\spanop\{\ket{K_5}\}.
 \label{loc:five-space}
\end{equation}
The image of $V_5$ therefore exhausts the entire allowed local sector.
The local product projector commutes with the synchronous permutation
projector $\Pi_5$, so it maps allowed inputs into this intersection and
satisfies
\begin{equation}
 \Pi_5(Q_{1/2}^{(5)})^{\otimes3}\Pi_5=P_5.
 \label{loc:local-compression5}
\end{equation}

Since $B_5=bI$ with $b=9/10$, the map $W_5=V_5/\sqrt b$ is an
isometric intertwiner. After sector decoding, Schur's lemma and
Eq.~\eqref{loc:five-space} give
\begin{equation}
 \mathcal S_{(1/2,1/2,1/2)}^{(5)} W_5\ket z
 =\ket z_{\mathrm{mag}}\otimes\ket{K_5}_{\mathrm{mult}}
 \label{loc:isometric-five}
\end{equation}
for a fixed unit vector $K_5$, including its phase. In the conventions
of Eqs.~\eqref{loc:cg-plus} and \eqref{loc:cg-minus}, the magnetic
spin-$1/2$ matrices are the original qubit matrices; hence the map on
$z$ is the identity, with its possible scalar phase absorbed in $K_5$.
Applying $P_5=W_5W_5^\dagger$ to $\ket\psi^{\otimes5}$ and using
$W_5^\dagger\ket\psi^{\otimes5}=D(\psi)\ket\psi/\sqrt b$ proves
Eq.~\eqref{loc:five-factorization} for every complex input.

For an explicit normalized tensor with this phase, put
$\ket G=(\ket{000}+\ket{111})/\sqrt2$ and use $D(G)=1/4$:
\begin{equation}
 \ket{K_5}
 =4\sqrt{\frac9{10}}
  (\bra G_{\mathrm{mag}}\otimes I)
  \mathcal S_{(1/2,1/2,1/2)}^{(5)}\ket G^{\otimes5}.
 \label{loc:k5-definition}
\end{equation}
The recursion above and this contraction specify all $125$ tensor
entries. They require no property of the unknown input. A global
phase change $\psi\mapsto e^{i\theta}\psi$ multiplies both sides of
Eq.~\eqref{loc:five-factorization} by $e^{5i\theta}$, as required.

At six copies, Eq.~\eqref{loc:intertwiner} maps the four residual
irreducible sectors of Eq.~\eqref{loc:residual} to their corresponding
local spin types. The three types $(1,0,0)$, $(0,1,0)$, and $(0,0,1)$
occur with multiplicity one by Eq.~\eqref{loc:g6}, so their full
allowed spin sectors belong to $\operatorname{ran}V_6$. The type
$(1,1,1)$ occurs twice, whereas the image of $V_6 R_{111}$ contains
only one copy. Thus its invariant multiplicity direction $K_D$ has a
single orthogonal complement $K_0$. This proves the completeness of
Eq.~\eqref{loc:six-factorization} throughout the allowed $(1,1,1)$
input sector, rather than only on selected inputs.

\subsection{Residual weights and physical branch probabilities}

Let $r_X=\Tr\rho_X^2$ and let $F_X$ swap the two residual qubits of
party $X$. The residual local triplet and singlet projectors are
$(I+F_X)/2$ and $(I-F_X)/2$. On $\ket\psi^{\otimes2}$,
\begin{align}
 \langle F_X\rangle&=r_X,&
 \langle F_XF_Y\rangle&=r_Z,\nonumber\\
 \langle F_AF_BF_C\rangle&=1,
 \label{loc:swap-identities}
\end{align}
where $X,Y,Z$ are distinct. The second identity uses equality of the
purities of complementary reductions of a pure state. Expanding the
projectors gives
\begin{align}
 \langle R_{111}\rangle
 &=\frac18\langle(I+F_A)(I+F_B)(I+F_C)\rangle
 =\frac{1+S}{4},\nonumber\\
 \langle R_{100}\rangle&=\frac{1+r_A-r_B-r_C}{4}.
 \label{loc:swap-weights}
\end{align}
Adding the last expression and its permutations gives
$\langle R_s\rangle=(3-S)/4$. The odd numbers of singlet factors
are absent because the full two-copy swap has eigenvalue $+1$.

For a residual sector $R_\nu$ with Gram eigenvalue $\lambda_\nu$,
the corresponding component of the certification projector is
\begin{equation}
 P_{6,\nu}=\lambda_\nu^{-1}V_6R_\nu V_6^\dagger.
 \label{loc:branch-projector}
\end{equation}
Consequently its probability on an identical-copy input is
\begin{equation}
 p_\nu=\frac{|D(\psi)|^2}{\lambda_\nu}
       \bra\psi^{\otimes2}R_\nu\ket\psi^{\otimes2}.
 \label{loc:branch-weight-proof}
\end{equation}
Inserting $\lambda_{111}=37/90$, $\lambda_s=1/2$, and the weights
above gives Eq.~\eqref{loc:branch-probabilities}. It also identifies
the accepted $K_D$ branch with $P_{6,111}$ and the three directly
accepted spin sectors with $P_{6,s}$.

\section{Conditional local discrimination}
\label{app:locc}

\subsection{Two fixed multiplicity states}

The bases and maps of Appendix~\ref{app:schur} specify the two
orthogonal real tensors needed by the six-copy protocol. Let
$\lambda_D=37/90$. Because $\ket{000}^{\otimes2}$ is the highest
weight vector in the residual $(1,1,1)$ representation,
\begin{equation}
 \ket{K_D}
 =\frac{(\bra{1,1,1}_{\mathrm{mag}}\otimes I)
       \mathcal S_{(1,1,1)}^{(6)}V_6\ket{000}^{\otimes2}}
       {\sqrt{\lambda_D}}.
 \label{loc:kd-definition}
\end{equation}
The bra denotes magnetic quantum numbers $m_A=m_B=m_C=1$.
Equation~\eqref{loc:b6} normalizes this tensor. The same
intertwining relation as in Eq.~\eqref{loc:intertwiner} shows that
all other magnetic components of this particular decoded vector
vanish.

To construct the orthogonal direction, take
$\ket W=(\ket{001}+\ket{010}+\ket{100})/\sqrt3$ and define
\begin{equation}
 \ket{K_0}
 =\sqrt{\frac{810}{37}}
 (\bra{1,1,1}_{\mathrm{mag}}\otimes I)
 \mathcal S_{(1,1,1)}^{(6)}\ket W^{\otimes6}.
 \label{loc:k0-definition}
\end{equation}
Each term of $\ket W^{\otimes6}$ has six down spins, giving total
magnetic quantum number $3$. Within three spin-one registers this
forces $m_A=m_B=m_C=1$. The normalization follows from
\begin{equation}
 \bigl\|\mathcal S_{(1,1,1)}^{(6)}\ket W^{\otimes6}\bigr\|^2
 =\frac{37}{810}.
 \label{loc:w-sector}
\end{equation}
The local projector is
\begin{equation}
 Q_1^{(6)}=
 \frac{\mathbf J^2(\mathbf J^2-6I)(\mathbf J^2-12I)}{80},
 \label{loc:q6}
\end{equation}
To evaluate Eq.~\eqref{loc:w-sector} explicitly, label a local
$m=1$ computational state by its two down-spin positions
$A\subset\{1,\ldots,6\}$. Let $B$ be the $15\times6$ incidence
matrix $B_{A,r}=1$ for $r\in A$ and zero otherwise. Its transpose
represents the raising operator from two down spins to one. Thus
the spin-one highest-weight projector in this magnetic sector is
$I-B(B^{\mathsf T}B)^{-1}B^{\mathsf T}$. Since
$B^{\mathsf T}B=4I+\boldsymbol1\boldsymbol1^{\mathsf T}$, its
matrix element depends only on $c=|A\cap A'|$:
\begin{equation}
 q_c=\delta_{c,2}-\frac c4+\frac1{10}.
 \label{loc:w-projector-elements}
\end{equation}
In particular, $q_0=1/10$, $q_1=-3/20$, and $q_2=3/5$.
Only terms assigning exactly two down spins to each party survive.
These are the $90$ ordered partitions of the six positions into
three pairs, each with amplitude $3^{-3}$. For any fixed partition,
sorting the $90$ possible second partitions by their three local
overlaps gives
\begin{align}
 \bigl\|\mathcal S_{(1,1,1)}^{(6)}\ket W^{\otimes6}\bigr\|^2
 =\frac{90}{3^6}\bigl(&q_2^3+12q_1^2q_2+16q_1^3\nonumber\\
       &+24q_0q_1^2+3q_0^2q_2\nonumber\\
       &+24q_0^2q_1+10q_0^3\bigr)
 =\frac{37}{810}.
 \label{loc:w-norm-proof}
\end{align}
Since $D(W)=0$, the vector
$\ket W^{\otimes6}$ is orthogonal to $\operatorname{ran}V_6$.
The spin projector preserves this orthogonality, and hence
$\braket{K_D}{K_0}=0$. The real Clebsch--Gordan coefficients and real
coefficients of $D$ make both tensors real. Equations
\eqref{loc:kd-definition} and \eqref{loc:k0-definition} fix their
phases and all $9^3$ entries before any unknown input is supplied.

\subsection{A zero-diagonal basis}

Perfect local discrimination of two orthogonal multipartite pure
states follows from Ref.~\cite{walgate2000}. A zero-diagonal basis
provides a direct construction~\cite{george2026}; for the real
tensors here it can be obtained using real orthogonal matrices alone.

Let $T$ be a real $d\times d$ matrix with $\Tr T=0$. There exists a
real orthogonal $Q$ satisfying
\begin{equation}
 (Q^{\mathsf T}TQ)_{aa}=0\quad\text{for every }a.
 \label{loc:zero-diagonal}
\end{equation}
To construct it, replace $T$ by its symmetric part
$H=(T+T^{\mathsf T})/2$, which has the same quadratic forms and
trace. If $H=0$, any real orthonormal basis works. Otherwise a
nonzero traceless symmetric matrix has eigenvalues
$\lambda_+>0$ and $\lambda_-<0$, with orthonormal eigenvectors
$e_+,e_-$. Set
\begin{equation}
 q=\sqrt{\frac{-\lambda_-}{\lambda_+-\lambda_-}}\,e_+
   +\sqrt{\frac{\lambda_+}{\lambda_+-\lambda_-}}\,e_-.
 \label{loc:zero-vector}
\end{equation}
Then $\|q\|=1$ and $q^{\mathsf T}Hq=0$. Choose an orthonormal
matrix $Z$ whose columns span $q^\perp$. The compression
$Z^{\mathsf T}HZ$ remains traceless because its trace is
$\Tr H-q^{\mathsf T}Hq=0$. Repeat on this $(d-1)$-dimensional
space. The resulting basis, beginning with $q$, satisfies
Eq.~\eqref{loc:zero-diagonal}. The recursion ends with a zero
one-dimensional compression. Eigenvalue degeneracies and vanishing
compressions allow arbitrary orthonormal choices, fixed once when
constructing the measurement.

\subsection{The sequential local measurement}

Write $u_{ijk}=\langle i,j,k|K_D\rangle$ and
$v_{ijk}=\langle i,j,k|K_0\rangle$. Both are real, normalized, and
orthogonal. Alice forms the transition matrix
\begin{equation}
 (T_A)_{ii'}=\sum_{jk}u_{ijk}v_{i'jk}
 \label{loc:alice-transition}
\end{equation}
and uses Eq.~\eqref{loc:zero-diagonal} to obtain an orthogonal
matrix $Q_A$. She measures in its column basis $\ket{q_a}$.
For result $a$, the conditional Bob--Charlie tensors are
\begin{equation}
 u^{(a)}_{jk}=\sum_i(Q_A)_{ia}u_{ijk},\qquad
 v^{(a)}_{jk}=\sum_i(Q_A)_{ia}v_{ijk}.
 \label{loc:alice-conditional}
\end{equation}
Their inner product is the $a$th diagonal entry of
$Q_A^{\mathsf T}T_AQ_A$, hence zero.

After receiving $a$, Bob forms
\begin{equation}
 (T_B^{(a)})_{jj'}=\sum_k u^{(a)}_{jk}v^{(a)}_{j'k}
 \label{loc:bob-transition}
\end{equation}
and measures in a zero-diagonal orthogonal basis
$Q_B^{(a)}$, whose columns are $\ket{q_{b|a}}$. Its trace vanishes
by the preceding conditional orthogonality. Charlie's two
unnormalized conditional vectors are
\begin{equation}
 x_{ab,k}=\sum_j(Q_B^{(a)})_{jb}u^{(a)}_{jk},\qquad
 y_{ab,k}=\sum_j(Q_B^{(a)})_{jb}v^{(a)}_{jk},
 \label{loc:charlie-vectors}
\end{equation}
and satisfy $\langle x_{ab}|y_{ab}\rangle=0$ for every $(a,b)$.
On receiving the two outcomes, Charlie uses the binary accept effect
\begin{equation}
 P_{ab}=\begin{cases}
 \ket{x_{ab}}\bra{x_{ab}}/\|x_{ab}\|^2,&x_{ab}\ne0,\\
 0,&x_{ab}=0.
 \end{cases}
 \label{loc:charlie-projector}
\end{equation}
Zero conditional vectors therefore cause no division and contribute
zero success probability. The complementary effect is $I-P_{ab}$.
This completes the instrument on every multiplicity input, including
states outside $\spanop\{K_D,K_0\}$.

The overall accept effect is explicitly
\begin{equation}
 E_D=\sum_{a,b}\ket{q_a}\bra{q_a}\otimes
          \ket{q_{b|a}}\bra{q_{b|a}}\otimes P_{ab}.
 \label{loc:locc-effect}
\end{equation}
The first two factors constitute a complete orthogonal resolution,
and $0\le P_{ab}\le I$, so $0\le E_D\le I$. Completeness of
Alice's and Bob's bases gives
\begin{align}
 \bra{K_D}E_D\ket{K_D}
 &=\sum_{ab}\|x_{ab}\|^2=1,\nonumber\\
 \bra{K_0}E_D\ket{K_0}&=0.
 \label{loc:perfect-discrimination}
\end{align}
Only the bases depend on the two fixed tensors; they are independent
of the input $\psi$. Classical communication follows the order
Alice, Bob, Charlie, with all quantum operations confined to each
party's own multiplicity register.

\subsection{Action on the full allowed input subspace}

Equation~\eqref{loc:perfect-discrimination} implies the vector
identities in Eq.~\eqref{loc:full-effect}: a zero expectation of a
positive operator implies membership in its kernel, applied to
$E_D$ for $K_0$ and to $I-E_D$ for $K_D$. For arbitrary magnetic
vectors $\ket x,\ket y$, even if they are not orthogonal, this gives
\begin{equation}
 (I\otimes E_D)(\ket x\ket{K_D}+\ket y\ket{K_0})
 =\ket x\ket{K_D}.
 \label{loc:correlated-effect}
\end{equation}
Thus the accepted probability is exactly $\|x\|^2$, with no
cross term. The same identity holds with an arbitrary additional
spectator system. It concerns the success effect; the local
measurements may record information about the accepted
multiplicity state.

For an explicit physical effect, let
$\mathcal S=\mathcal S_{(1,1,1)}^{(6)}$. Then
\begin{align}
 L_6={}&Q_1^{(6)}\otimes Q_0^{(6)}\otimes Q_0^{(6)}
       +\text{two permutations}\nonumber\\
     &+\mathcal S^\dagger(I_{\mathrm{mag}}\otimes E_D)\mathcal S,
 \label{loc:physical-effect6}
\end{align}
where
\begin{equation}
 Q_0^{(6)}=-\frac{(\mathbf J^2-2I)(\mathbf J^2-6I)
                         (\mathbf J^2-12I)}{144}.
 \label{loc:q60}
\end{equation}
The four spin branches are mutually orthogonal, so $L_6$ is a
valid effect. The sector identification in Appendix~\ref{app:schur}
and Eq.~\eqref{loc:correlated-effect} prove
\begin{equation}
 \Pi_6L_6\Pi_6=P_6.
 \label{loc:local-compression6}
\end{equation}
Multiplying only the last term of Eq.~\eqref{loc:physical-effect6}
by $37/45$ gives the optimal binary-readout effect. In the residual
representation this replacement changes the two coefficients of
$V_6R_\nu V_6^\dagger$ to the common value $2$:
$(37/45)/(37/90)=1/(1/2)=2$. Its compression is therefore
$2V_6V_6^\dagger$, which attains the spectral bound and has
probability $t/8$.

\section{Statistical inference and comparison methods}
\label{app:statistics}

\subsection{Intervals from a binary event}

For $X=x$ successes in $N$ independent batches, the equal-tailed
Clopper--Pearson interval~\cite{clopper1934} with error probability $\alpha$
has endpoints
\begin{align}
 p_L&=\begin{cases}
 0,&x=0,\\
 B^{-1}_{\alpha/2}(x,N-x+1),&x>0,
 \end{cases}\label{eq:cp-lower}\\
 p_U&=\begin{cases}
 1,&x=N,\\
 B^{-1}_{1-\alpha/2}(x+1,N-x),&x<N,
 \end{cases}\label{eq:cp-upper}
\end{align}
where $B^{-1}_q(a,b)$ is a beta-distribution quantile. These endpoints
follow by inverting the two binomial tails, each with error at most
$\alpha/2$. For $p=\kappa t$, intersect $[p_L,p_U]$ with $[0,\kappa]$
and divide by $\kappa$ to obtain a confidence set for $t$. Taking the
square roots of its endpoints gives the corresponding set for $\tau$.
An empty physical intersection remains an empty confidence set.

Zero successes give a one-sided upper bound
$p\le1-\alpha^{1/N}$ at confidence $1-\alpha$. Thus the absence of a
certificate constrains the invariant but does not identify the algebraic
null set exactly at finite $N$.

\subsection{The six-copy branch record}

Let $(X_D,X_s,X_f)$ have a multinomial distribution with probabilities
$(p_D,p_s,1-p_D-p_s)$. The marginals are binomial, but
$\operatorname{Cov}(X_D,X_s)=-Np_Dp_s$. Substitution into
Eq.~\eqref{eq:weighted-estimate} gives Eq.~\eqref{eq:weighted-variance}.
For a binary event produced by thinning, let $Z_i$ be the accepted bit in
batch $i$, and let $R_i$ be its unthinned branch. Then
\begin{equation}
 \mathbb E[Z_i\mid R_i]
 =a\,1_{\{R_i=D\}}+1_{\{R_i=s\}}.
\end{equation}
The law of total variance gives Eq.~\eqref{eq:variance-gap} because only
the $D$ branch has conditional variance $a(1-a)$.

To construct a simultaneous interval, form binomial intervals
$[l_D,u_D]$ and $[l_s,u_s]$, each with error probability $\alpha/2$.
The union bound gives
\begin{equation}
 t\in[8(al_D+l_s),\,8(au_D+u_s)]\cap[0,1]
 \label{eq:weighted-ci}
\end{equation}
with coverage at least $1-\alpha$. This construction uses integer counts
and does not require their independence. Its conservatism can be reduced
by optimizing the linear functional $8(ap_D+p_s)$ over a joint multinomial
confidence region.

\subsection{Noisy certification with calibration}

Suppose a calibration independent of the test data gives
$p_{\rm null}\le p_0$ except with probability $\alpha_{\rm cal}$.
For a fixed number $N$ of test batches, reject the null when
\begin{equation}
 \sum_{k=x}^N\binom Nk p_0^k(1-p_0)^{N-k}
 \le\alpha_{\rm test}.
 \label{eq:calibrated-test}
\end{equation}
Monotonicity of the binomial tail and a union bound give total type-I
error at most $\alpha_{\rm cal}+\alpha_{\rm test}$.
For a specified finite collection of $K$ null sources, simultaneous
one-sided calibration bounds can be obtained by assigning error
$\alpha_{\rm cal}/K$ to each source and taking the largest bound.
The resulting guarantee covers that collection. A larger source class
requires a bound over the entire class, for example from
Eq.~\eqref{eq:robust-bound}.

Calibration and testing both consume copies. If $N_j^{\rm cal}$ batches
are allocated to null source $j$, their input cost is
$n\sum_jN_j^{\rm cal}$. At fixed calibration size the null bound does not
shrink when the test size increases. Allocating copies between calibration
and testing is therefore part of the statistical design.

\subsection{Numerical estimation comparisons}

The simulated event measurements use binomial or multinomial outcome counts
at their ideal instrument probabilities. Figure~\ref{fig:statistics} uses
1000 independent repetitions for each fixed state, protocol, and input
budget. A sample variance is compared with the analytic variance of the
same untruncated estimator.
For $R$ repetitions, the variance of the unbiased sample variance is
\begin{equation}
 \operatorname{Var}(s^2)=\frac1R\left[
 \mu_4-\frac{R-3}{R-1}v^2\right],
 \label{eq:variance-mc-error}
\end{equation}
where $v$ and $\mu_4$ are the variance and fourth central moment of one
estimate. The error bars use the square root of this expression, with
moments obtained from the binomial or weighted categorical distribution.

For tomography, the budget is distributed as evenly as possible among the
27 settings in $\{X,Y,Z\}^{\otimes3}$. Compatible settings are pooled to
estimate the Pauli expectations. Their linear inversion gives
\begin{equation}
 \widehat\rho_{\rm lin}=\frac18
 \sum_{P\in\{I,X,Y,Z\}^{\otimes3}}\widehat{\langle P\rangle}P.
\end{equation}
The eigenvector with largest eigenvalue provides the pure-state estimate,
from which $t$ is evaluated. This uses the same pure-source assumption as
the event protocols.

The canonical-form baseline uses the three local unitaries and objective
$P_{001}+P_{010}+P_{011}$ of Ref.~\cite{perezsalinas2020}. A Powell search
receives sampled counts for each objective evaluation, with up to three
initializations sharing a training allocation of 70\% of the total
budget. Final independent counts give the product statistic
$\widehat\tau=4\widehat P_{000}\widehat P_{111}$. We retain every fit when
computing mean squared errors. A separate binomial interval for the
residual objective tests whether its upper 95\% endpoint is below $0.01$.
This interval concerns the residual population, not the error in the
estimated tangle.

Table~\ref{tab:mse-example} gives an equal-input-budget example, including
one Monte Carlo standard error of each reported mean squared error.
For squared errors $e_r=(\widehat t_r-t)^2$, that standard error is
$s_e/\sqrt R$, where $R$ is the number of independent repetitions.
For this input the canonical-form implementation used 43 copies per objective
evaluation and 162 evaluations across three initializations: 6966 training
copies and 3034 final readout copies per repetition. None of its final
readouts established a residual objective
below $0.01$ at the stated confidence. The input amplitudes and all
outcome counts accompany the numerical data.

\begin{table}[t]
\caption{Estimation of $t$ for one complex pure state with $t=0.49931$
and $S=1.68521$, at a budget of $10^4$ input copies. Parentheses give one
Monte Carlo standard error in the last quoted digits. The event rows use
1000 repetitions; the two reconstruction rows use 30.}
\label{tab:mse-example}
\begin{ruledtabular}
\begin{tabular}{l c}
Method & Mean squared error\\\hline
Direct four-copy circuit & $0.0476(23)$\\
Complete four-copy singlet & $0.00982(42)$\\
Five-copy retention & $0.00347(15)$\\
Six-copy binary & $0.002270(97)$\\
Six-copy weighted & $0.001984(85)$\\
Pure-state Pauli tomography & $0.00048(10)$\\
Sampled canonical-form fit & $0.173(14)$
\end{tabular}
\end{ruledtabular}
\end{table}

\section{Circuit implementation and preparation errors}
\label{app:implementation}

\subsection{Registers and circuit resources}

Each party applies its local Schur transform before reading a coarse spin
flag. At four copies an auxiliary qubit records whether $J=0$. At five
copies it records whether $J=1/2$; the two-dimensional magnetic register
remains coherent and supplies that party's retained qubit. These protocols
use, respectively, 12 and 15 input qubits and three auxiliary qubits in
total. Each party measures only its spin flag and communicates its result.

At six copies the local encoding reserves four qubits for the multiplicity
label and two for the magnetic label, together with two ancillary spin
flags. The three parties therefore use 18 input qubits and six ancillary
qubits. The $(1,1,1)$ sector proceeds through Alice's measurement, Bob's
conditional measurement, and Charlie's conditional acceptance. Unused
basis states in the encoded registers are part of the physical circuit
space, including in calculations with gate errors.

We synthesize the local unitaries using arbitrary single-qubit rotations
and CX gates. The accepted binary labels are grouped into disjoint
subcubes before coupling them to the spin flags, requiring one, two, and
six controlled flips per party at four, five, and six copies, respectively.
A second construction resolves the Clebsch--Gordan steps
into two-level rotations and explicit label permutations. The latter
construction uses more CX gates at these small sizes; the resource
comparison in the main text uses the first synthesis. Linear-connectivity
counts include swaps and the recovery of the designated output wire order.
Charlie's final rank-one acceptance test applies the inverse preparation
of its normalized acceptance vector, followed by a computational-basis
measurement that accepts the all-zero outcome. This implements the same
acceptance effect on all 16 multiplicity-register basis states.
With one additional clean work qubit per party, the six-copy spin filter
can instead be compiled with 1899 all-to-all CX gates per party; the work
qubit returns to zero for every ideal input. The main resource comparison
uses the implementation without this additional qubit.
Input-state preparation is excluded in both constructions and in the shallow
reference circuit. Initial filters act in parallel across the three
parties. The subsequent conditional measurements are sequential; the
reported quantum depth excludes communication latency.

For a protocol whose first-success probability per batch is $p$, let
$R$ be its expected total gate cost until success. Conditioning on the
first batch gives $R=\mathbb E[G]+(1-p)R$, hence
$R=\mathbb E[G]/p$. This remains true when a batch's gate cost and its
success indicator are correlated. Average and worst-case path costs are
therefore different resource measures.

\subsection{Bounds for imperfect inputs and effects}

For any effect $0\le M\le I$,
$|\Tr M(\rho-\sigma)|\le T(\rho,\sigma)$.
A telescoping expansion of the product states gives
\begin{equation}
 T\!\left(\bigotimes_r\rho_r,\sigma^{\otimes n}\right)
 \le\sum_r T(\rho_r,\sigma).
\end{equation}
If $\sigma=\ket\phi\bra\phi$ is a common null input, the ideal success
probability on $\sigma^{\otimes n}$ vanishes. Adding
$|\Tr(\widetilde M-M)\rho|\le\epsilon_M$ proves
Eq.~\eqref{eq:robust-bound}.

For an identical pure input $\psi$ close to a null state $\phi$, positivity
implies $M\ket\phi^{\otimes n}=0$. Since $M\le I$, its expectation on
$\ket\psi^{\otimes n}$ is at most
\begin{equation}
 1-|\braket{\phi}{\psi}|^{2n}
 \le1-(1-\eta^2)^n\le n\eta^2.
\end{equation}
The quadratic dependence uses the coherent pure-state perturbation and
is distinct from the general product-state bound.

For the success map, set
$\Delta=\widetilde\sigma-p\ket\psi\bra\psi$.
Then $|\widetilde p-p|=|\Tr\Delta|\le\epsilon_s$, and
\begin{align}
 \left\|\widetilde\sigma-
       \widetilde p\ket\psi\bra\psi\right\|_1
 &\le\|\Delta\|_1+|p-\widetilde p|\\
 &\le2\epsilon_s.
\end{align}
Dividing by $2\widetilde p$ proves
Eq.~\eqref{eq:retention-robustness}. In particular, a bound on the success
effect alone cannot fix the postmeasurement quantum state.

A universal mixed-state version of a conclusive positive event is
excluded by a full-rank null input. If a positive effect accepts the
maximally mixed, fully separable state with zero probability, then
\begin{equation}
 0=\Tr[M(I/8)^{\otimes n}]=8^{-n}\Tr M
 \quad\Longrightarrow\quad M=0.
\end{equation}
Mixed-state polynomial witnesses instead use signs or inequalities of
Hermitian expectation values~\cite{fraser2026}. Finite-error certification
can also use the calibrated test of Appendix~\ref{app:statistics}.

\subsection{Physical noise models}

Preparation noise is applied before the local instrument. The source
models considered are independent per-copy mixing,
$(1-\epsilon)\ket\psi\bra\psi+\epsilon I/8$; local phase flips;
amplitude damping; and copy-dependent local rotations. A phase-flip
channel with probability $q$ is
$(1-q)\rho+qZ\rho Z$. Amplitude damping uses
\begin{equation}
 E_0=\ket0\bra0+\sqrt{1-\gamma}\ket1\bra1,\qquad
 E_1=\sqrt\gamma\ket0\bra1.
\end{equation}
The single-qubit channels act independently at the parties and copies.
A rotation drift applies $\exp(-i\vartheta\mathbf u\cdot\boldsymbol\sigma/2)$,
with $\mathbf u$ uniform on the unit sphere and $\vartheta$ a zero-mean
Gaussian angle. The angular standard deviation is specified in radians;
axes and angles are sampled independently for each party and copy.
All models act in the full physical tensor-product space.

For the circuit-level channel, each CX gate is followed by $I$ with
probability $1-p_{\rm CX}$ and by each other two-qubit Pauli with probability
$p_{\rm CX}/15$. Single-qubit gates are ideal in this model. Conditional
six-copy gates are applied on their actual measurement branches. A
stochastic trajectory supplies a full instrument acceptance probability;
the trajectory mean and its standard error describe numerical integration
of the noise channel, not finite-shot uncertainty of an experimental
frequency. Conditional retained states are formed by averaging
unnormalized successful outputs and then dividing by their trace.

\ifwithbibliography
\bibliography{references}
\fi

\end{document}